\documentclass[aps,prd,preprint,superscriptaddress,tightenlines,nofootinbib]{revtex4}

\usepackage[final]{graphicx}
\usepackage{dcolumn}
\usepackage{bm}

\def\cleoc{CLEO-c}

\def\ep{e^+}
\def\em{e^-}

\def\mup{\mu^+}
\def\mum{\mu^-}

\def\pip{\pi^+}
\def\pim{\pi^-}
\def\piz{\pi^0}
\def\pipm{\pi^\pm}

\def\Kp{K^+}

\def\Kz{K^0}

\def\KS{K^0_S}

\def\Kpm{K^\pm}

\def\Dp{D^+}
\def\Dm{D^-}
\def\Dz{D^0}
\def\Dbar{\bar{D}}
\def\Dzbar{\bar{D}^0}

\def\Dsp{D_s^+}
\def\Dsm{D_s^-}

\def\Jpsi{J/\psi}
\def\psiprime{\psi'}
\def\psidprime{\psi(3770)}

\def\Dzkpi{\Dz\to K^-\pi^+}
\def\Dzkpipiz{\Dz\to K^-\pi^+\pi^0}
\def\Dzkpipipi{\Dz\to K^-\pi^+\pi^+\pi^-}
\def\Dpkpipi{\Dp\to K^-\pi^+\pi^+}
\def\Dpkspi{\Dp\to \KS\pi^+}
\def\Dzbarkpi{\Dzbar\to K^+\pi^-}
\def\Dzbarkpipiz{\Dzbar\to K^+\pi^-\pi^0}
\def\Dzbarkpipipi{\Dzbar\to K^+\pi^-\pi^-\pi^+}
\def\Dmkpipi{\Dm\to K^+\pi^-\pi^-}
\def\Dmkspi{\Dm\to \KS\pi^-}

\def\BDzkpi{\calB(\Dzkpi)}
\def\BDzkpipiz{\calB(\Dzkpipiz)}
\def\BDzkpipipi{\calB(\Dzkpipipi)}
\def\BDpkpipi{\calB(\Dpkpipi)}
\def\BDpkspi{\calB(\Dpkspi)}

\def\Ecm{E_\mathrm{cm}}
\def\Ez{E_0}

\def\Gevc{GeV/$c$}
\def\Gevcsq{GeV/$c^2$}

\def\Mevcsq{MeV/$c^2$}

\def\Mmsq{MM^2}

\def\DeltaE{\Delta E}
\def\ED{E(D)}
\def\pD{p(D)}
\def\MD{M(D)}
\def\MDbar{M(\Dbar)}
\def\MDz{M_{\Dz}}
\def\MDp{M_{\Dp}}

\def\mbar{\bar{m}}
\def\mavg{\hat{m}}
\def\deltam{\delta m}

\def\xim{\xi_m}
\def\sigmab{\sigma_b}
\def\sigmac{\sigma_c}
\def\sigmadelta{\sigma_\delta}

\def\MDavg{\widehat{M}(D)}

\def\sigmaL{\sigma_L}
\def\sigmaR{\sigma_R}
\def\rA{\rho_A}
\def\rL{\rho_L}
\def\rR{\rho_R}

\def\calB{{\cal B}}
\def\Lum {{\cal L}}
\def\pbinv{~pb$^{-1}$}

\def\Bi{\calB_i}
\def\Ni{N_i}
\def\effi{\epsilon_i}
\def\Bj{\calB_j}
\def\Nj{N_j}
\def\effj{\epsilon_j}
\def\Nij{N_{ij}}
\def\effij{\epsilon_{ij}}
\def\Nii{N_{ii}}
\def\effii{\epsilon_{ii}}

\def\NDDbar{N_{D\Dbar}}
\def\NDzDzbar{N_{\Dz\Dzbar}}
\def\NDpDm{N_{\Dp\Dm}}

\def\eg{{\it e.g.}}
\def\ie{{\it i.e.}}
\def\vs{{\it vs.}}
\def\etal{{\it et al.}}

\catcode`\@=11 
\newcommand{\Begitem}{\begin{list}
{\csname\@itemitem\endcsname}
{\ifnum \@itemdepth >3 \@toodeep\else \advance\@itemdepth 1 
\edef\@itemitem{labelitem\romannumeral\the\@itemdepth} 
  \parsep  2pt plus 1pt minus 1pt 
  \parskip 0pt plus 1pt minus 1pt 
  \topsep  0pt plus 1pt minus 1pt             
  \itemsep 0pt plus 1pt minus 1pt \fi}}
\newcommand{\Enditem}{\end{list}}
\catcode`\@=12 

\def\Fig#1{Fig.~\ref{#1}}
\def\Sec#1{Sec.~\ref{#1}}
\def\Tab#1{Table~\ref{#1}}

\def\Begfigure#1{\begin{figure}[#1]\begin{center}}
\def\Endfigure#1#2{\caption{#1}\label{#2}
               \end{center}\end{figure}}

\def\Begtable#1#2#3{\begin{table}[#1]
\caption{#2}\vspace*{-1.5ex}\label{#3}\begin{center}}
\def\Endtable{\end{center}\end{table}}

\newcommand{\Arraystretch}{1.2}

\newcommand{\Begtabular}[1]{\renewcommand{\arraystretch}{\Arraystretch}
                         \begin{center}\begin{tabular}{#1}\hline\hline}
\newcommand{\Endtabular}{\hline\hline    \end{tabular}\end{center}}

\begin{document}

\preprint{\parbox{1.5in}{CLEO CONF 04-10\\ ICHEP04 ABS11-0775}}  

\title{\boldmath Hadronic Branching Fractions of $D^0$ and $D^+$, and\\ 
$\sigma(e^+e^- \to D\Dbar$) at $\Ecm = 3.77$~GeV}
\thanks{Submitted to the 32$^{\rm nd}$ International Conference on High
Energy Physics, Aug 2004, Beijing}


\author{B.~I.~Eisenstein}
\author{G.~D.~Gollin}
\author{I.~Karliner}
\author{D.~Kim}
\author{N.~Lowrey}
\author{P.~Naik}
\author{C.~Sedlack}
\author{M.~Selen}
\author{J.~J.~Thaler}
\author{J.~Williams}
\author{J.~Wiss}
\affiliation{University of Illinois, Urbana-Champaign, Illinois 61801}
% ...
\collaboration{CLEO Collaboration} 
\noaffiliation



\date{\today}

\begin{abstract} 
Using nearly 60\pbinv\ of data collected with the \cleoc\ detector at the
$\psidprime$ resonance, we measure absolute branching fractions for three
$\Dz$ and two $\Dp$ Cabibbo-allowed hadronic decay modes, and the cross
section for 
$\ep\em \to D\Dbar$ at $\Ecm = 3.77$~GeV.  We report
preliminary measurements of the reference branching fractions, 
$\calB(\Dzkpi) = (3.92\pm 0.08\pm 0.23)\%$ and 
$\calB(\Dpkpipi) = (9.8\pm 0.4\pm 0.8)\%$, and  
preliminary measurements of other major branching fractions,
$\calB(\Dzkpipiz)  = (14.3\pm 0.3\pm 1.0)\%$, 
$\calB(\Dzkpipipi) = (8.1\pm 0.2\pm 0.9)\%$, and
$\calB(\Dpkspi)    = (1.61\pm 0.08\pm 0.15)\%$.
We determine preliminary values of the cross sections, 
$\sigma(\ep\em\to \Dz\Dzbar) = (3.47\pm 0.07\pm 0.15)$~nb,
$\sigma(\ep\em\to \Dp\Dm)    = (2.59\pm 0.11\pm 0.11)$~nb, and
$\sigma(\ep\em\to D\bar D )  = (6.06\pm 0.13\pm 0.22)$~nb.
We note that the Monte Carlo simulations used in calculating efficiencies in this
analysis included final state radiation.  However, the branching fractions used
in the Particle Data Group averages do not include this effect.  If we had not
included final state radiation in our simulations, branching fractions would have
been 0.5\% to 2\% lower.  

\end{abstract}

\maketitle

\section{Introduction}

We present preliminary absolute measurements of the Cabibbo-allowed $\Dz$ and
$\Dp$ branching fractions, $\BDzkpi$, $\BDzkpipiz$,
$\BDzkpipipi$, $\BDpkpipi$, and $\BDpkspi$.  Two of these branching
fractions, $\BDzkpi$ and $\BDpkpipi$, along with the branching fraction for
$\Dsp\to\phi\pip$, are particularly important because essentially all other
$\Dz$, $\Dp$, and $\Dsp$ branching fractions are determined from ratios  to
one or the other of these branching fractions~\cite{PDG}.  As a result, nearly
all branching fractions in the weak decay of heavy quarks are ultimately tied
to one of these three branching fractions, called reference branching fractions
in this paper.  Furthermore, these reference branching fractions appear in
many measurements of CKM matrix elements for $c$ and $b$ quark decay.

We note that the Monte Carlo simulations used in calculating efficiencies in this
analysis included final state radiation.  However, the branching fractions used
in the Particle Data Group averages do not include this effect.  If we had not
included final state radiation in our simulations, the branching fractions we
measure would have been 0.5\% to 2\% lower than they are.  

\section{Branching Fractions and Production Cross Sections}

The data for these measurements were obtained in $\ep\em$ collisions at 
$\Ecm = 3.77$~GeV, the peak of the $\psidprime$ resonance.  At this energy, no
additional hadrons accompany the
$\Dz\Dzbar$ and $\Dp\Dm$ pairs that are produced.  These unique 
$D\Dbar$ final states provide a powerful tool for
addressing the most vexing problem in measuring absolute $D$ branching
fractions at higher energies -- the difficulty of accurately determining the
number of $D$ mesons produced.  Following the MARK~III
collaboration~\cite{markiii-1,markiii-2}, reconstruction of one $D$ meson (called
single tag or ST) serves to tag the event as either $\Dz\Dzbar$ or $\Dp\Dm$. 
Branching fractions for
$\Dz$ or $\Dp$ decay can then be obtained from the number of ``double
tag'' (DT) events in which both the $D$ and the $\Dbar$ are reconstructed,
without knowledge of luminosity or the total number of
$D\Dbar$ events produced.  Also, since $\Ecm$ is
below threshold for production of $\Dsp\Dsm$, accurate absolute measurement of
$\calB(\Dsp\to\phi\pip)$ using the techniques in this paper remains a challenge
for the future. 

If no distinctions are made between $D$ and $\Dbar$ decays and their
efficiencies, the number of ST events $\Ni$ observed in the decay mode
$i$ (and its charge conjugate) with branching fraction $\Bi$ and detection
efficiency $\effi$ will be,
\[
\Ni = 2 \NDDbar \Bi \effi,
\]
where $\NDDbar$ is the number of $D\Dbar$ events produced in the
experiment.  Then, the number of DT events with $D\Dbar$ pairs reconstructed
in modes $i$ and
$j$ will be,
\[ 
\Nij = 2 \NDDbar \Bi\Bj \effij ~~(\mathrm{if}~ i \neq j) \mathrm{~~or~~}
\Nii =   \NDDbar \Bi^2 \effii  ~~(\mathrm{if}~ i = j) 
\]
Hence, the ratios of DT events ($\Nij$) to ST events
($\Nj$) provide absolute measurements of the branching fraction
$\Bi$,
\[ 
\Bi = {\Nij \over \Nj}{\effj \over \effij} \mathrm{~~or~~} 
\Bi = {2 \Nii \over \Ni}{\effi \over \effii}.
\] 
Note that $\effij \approx \effi\effj$, so $\effj$ nearly cancels from the ratio
$\effj/\effij$.  Hence, a branching fraction $\Bi$ obtained using this procedure
is nearly independent of the efficiencies of tagging
modes. Of course, $\Bi$ is sensitive to $\effi$ and its uncertainties.

Estimating errors and combining measurements using just these
expressions is very difficult because $\Nij$ and $\Nj$ are correlated (whether
or not $i=j$) and measurements of $\Bi$ using different tagging modes $j$ are
also correlated.  
To address this problem, we have developed a $\chi^2$ fitting
procedure~\cite{brfit} which fits simultaneously all charged and neutral $D$
branching fractions and the numbers of charged and neutral $D\Dbar$ pairs. 
Although the $\Dz$ and $\Dp$ branching fractions are statistically independent,
systematic effects  introduce significant correlations among them.  Therefore,
we fit both charged and neutral $D$ mesons simultaneously, and we include both
statistical and systematic uncertainties, as well as their correlations, in
the fit.  We also perform corrections for backgrounds, efficiency, and crossfeed
among modes directly in the fit, as the sizes of these adjustments depend on
the fit parameters.  Thus, all experimental measurements, such as event
yields, efficiencies, and background branching fractions, are treated in a
consistent manner.  We actually fit $D$ and $\Dbar$ yields separately in
order to accommodate possible differences in efficiency, but charge conjugate
branching fractions are constrained to be equal.

The number of $D\Dbar$ events that were produced can be calculated from
\[
\NDDbar = {\Ni \Nj \over 2 \Nij} {\effij \over \effi \effj} \mathrm{~~or~~}
\NDDbar = {\Ni^2 \over 4 \Nii} {\effii \over \effi^2}.
\]
The production cross sections for $\Dz\Dzbar$ and $\Dp\Dm$ can then be
obtained by combining $\NDzDzbar$ and $\NDpDm$ -- which are determined in
the branching fraction fit -- with the integrated luminosity $\Lum$. 
Note that $\NDDbar$ obtained by this procedure is almost independent of
the efficiencies.

Charge conjugate particles and decay modes are always implied in this
paper unless stated otherwise. 

\section{The \cleoc\ Detector}

The \cleoc\ detector is a modification of the CLEO~III
detector~\cite{cleoiidetector,cleoiiidr,cleorich} in which the silicon-strip
vertex detector was replaced with a six-layer vertex drift chamber, whose
wires are all at small stereo angles to the beam axis~\cite{cleocyb}.  The
charged particle tracking system, consisting of the vertex drift chamber and
a 47-layer central drift chamber~\cite{cleoiiidr} operates in a 1.0~T
magnetic field, whose direction is along the beam axis. The momentum resolution
achieved with the tracking system is $\sim 0.6$\% at
$p=1$~\Gevc. Photons are detected in an electromagnetic calorimeter
consisting of about 8000 CsI(Tl) crystals~\cite{cleoiidetector}.  The
calorimeter attains a photon  energy resolution of 2.2\% at $E_\gamma=1$~GeV
and 5\% at 100~MeV. The solid angle coverage for charged and neutral
particles of the
\cleoc\ detector is 93\% of $4\pi$.  We utilize two devices to obtain particle
identification information to separate $K^\pm$ from $\pi^\pm$: 
the central drift chamber (which provides measurements of ionization energy loss
-- $dE/dx$) and a cylindrical ring-imaging Cherenkov (RICH)
detector~\cite{cleorich} surrounding the central drift chamber.   The 
solid angle of the RICH
detector for separation of $\Kpm$ from $\pipm$ is 80\% of $4\pi$.   
The combined
$dE/dx$-RICH particle identification procedure has a pion or kaon efficiency
$>90$\% and a probability of pions faking kaons (or vice versa)
$<5$\%.

The integrated luminosity ($\cal{L}$) of the data sets was measured 
using $\gamma\gamma$ events in the calorimeter~\cite{lumins}. Event counts
were normalized with a Monte Carlo (MC) simulation based on the
Babayaga~\cite{bby} event  generator combined with GEANT-based~\cite{geant}
detector modeling.

\section{Data Sample and Event Selection}
\label{sec-data&cuts}

In this analysis, we utilized a total integrated luminosity of $\Lum = 57.2 \pm
1.7$~\pbinv\ of
$\ep\em$ data collected at $\Ecm = 3.77$~GeV.
The data were produced with the Cornell Electron Storage
Ring (CESR) operating in a new configuration~\cite{cleocyb} that includes six
wiggler magnets to enhance synchrotron radiation damping at energies in the charm
threshold region.  Since these data were obtained, six additional wiggler magnets
have been installed in CESR and future CESR operation in this energy region will
be with the full complement of twelve wiggler magnets.  The spread in $\Ecm$ with
the six wiggler magnets is 2.3~MeV. 

We reconstructed all events that satisfied the \cleoc\ trigger and preserved
them for further analyses.  Most \cleoc\ analyses of $D$ meson decays utilize a
subset of the data in which each event contains at least one $D$ meson candidate
selected by a standard set of requirements.  This selection procedure begins with
standardized requirements for $\pipm$, $\Kpm$, $\piz$, and $\KS$ candidates.  

Charged track candidates must pass the
following quality requirements: the momentum of the track $p$ must be in the
range $0.050 \leq p \leq 2.0$~\Gevc, the track must pass within 0.5~cm of the
origin in the $x$-$y$ plane (transverse to the beam direction),
the track must pass within 5.0~cm of the origin in the $z$
direction, the polar angle $\theta$ was required to be in the range
$|\cos\theta| < 0.93$, and the number of hits reconstructed on the track was
required to be at least half of the number of layers traversed by the track.
Both requirements on the distance of the track from the origin of the
coordinate system (which is close to the point where the two beams intersect) are
approximately five times the standard deviation for the corresponding measurement.

We identified charged track candidates as pions or kaons using $dE/dx$ and
RICH information.  Each track was considered as both 
a potential $K^\pm$ and $\pi^\pm$ candidate.  
In the rare case that no useful information of either
sort was available, we did not change this assignment.

If $dE/dx$ information was available, we calculate 
$\chi^2_E(\pi)$ and $\chi^2_E(K)$ from the $dE/dx$
measurements, the expected $dE/dx$ for pions and kaons of that momentum, 
and the appropriate standard deviations.  
We rejected tracks as kaon candidates when $\chi_E^2(K)$ was greater than 9, 
and similarly for pions.  
The difference $\chi^2_E(\pi) - \chi^2_E(K)$ was also calculated.
If $dE/dx$ information was not
available this $\chi^2$ difference was set equal to 0.

We used RICH information if the track momentum was sufficiently above the $K$
threshold  ($p > 0.55$~\Gevc), and the track was within the RICH acceptance
($|\cos\theta| < 0.8$).  We then rejected tracks as kaon candidates 
when the number of Cherenkov photons detected 
for the the kaon hypothesis was less than three, and similarly for pions.  
When there were at least three photons for the hypothesis under consideration, 
and information was available for both pion and kaon hypotheses, 
we calculated a $\chi^2$ difference for the RICH, 
$\chi^2_R(\pi) - \chi^2_R(K)$, 
from the locations of Cherenkov photons and the track parameters.  
Otherwise, we set this $\chi^2$ difference equal to 0. 

Then we combined these $\chi^2$ differences in an overall
$\chi^2$ difference,
$\Delta \chi^2 \equiv \chi^2_E(\pi) - \chi^2_E(K) + 
\chi^2_R(\pi) - \chi^2_R(K)$.  
If $\Delta \chi^2 \le 0$, we used the track as a pion candidate; 
If $\Delta \chi^2 \ge 0$, we used the track as a kaon candidate.  

We formed neutral pion candidates from pairs of photons, each of
whose energy was greater than 30~MeV and whose showers pass photon quality
requirements.  An unconstrained mass $M(\gamma\gamma)$ was
calculated from the energies and momenta of the two photons.  This mass was
required to be within $3\sigma$ with a nominal $\piz$ mass value which varied
slightly with the total momentum of the $\piz$ candidate.  The 
energy and momentum obtained from a kinematic fit of the two photon
candidates to the mass $M_{\piz}$ from the PDG~\cite{PDG} was then used in
further analysis.  
  
We built $\KS$ candidates from pairs of tracks that intersect a vertex.  We
then subjected the tracks to a constrained vertex fit and used the
resulting track parameters to calculate the invariant mass, $M(\pip\pim)$. 
We called the track pair a $\KS$ candidate if the invariant mass
$M(\pip\pim)$ was within $3\sigma$, with $\sigma = 4$~\Mevcsq, of the mass
$M_{\Kz}$ from the PDG~\cite{PDG}.

After event reconstruction, the next major step was selecting
$D\Dbar$ events with relatively loose requirements for further analysis.  
In this \cleoc\ standard selection process we reconstructed $\Dz$ candidates in 18
important decay modes and $\Dp$ candidates in 6 important decay modes, although
only three $\Dz$ modes and two $\Dp$ modes were used in this analysis. Two
requirements were placed on $\Dz$ and $\Dp$ candidates formed from
$\pipm$, $\Kpm$, $\piz$ and $\KS$ candidates selected using the 
requirements described above.
We calculate the energy difference,
$\DeltaE \equiv \ED - \Ez$, where
$\ED$ is the total energy of the particles in the $D$ candidate and $\Ez$
is the energy of the $e^\pm$ beams.  We accept candidates for further
analysis if $|\DeltaE| < 100$~MeV. We calculate the mass of the $D$ candidate
$\MD$ by substituting the beam energy $\Ez$ for the energy $E(D)$
of the $D$ candidate, \ie, 
$M^2(D)c^4 \equiv \Ez^2 - p^2(D)c^2$, where $\pD$ is the total momentum of
the particles in the $D$ candidate.  Then we required $\MD > 1.83$~\Gevcsq\
to retain a $D$ candidate for further analysis.  We included all events
with $D$ candidates in any of the 24 decay modes that satisfied these
loose requirements.  
 
For the analysis described in this paper, we restricted our attention to the
five decay modes mentioned previously, and we refined the requirements for
acceptable $D$ candidates that were described in the previous
paragraph.  First, we used the more restrictive requirements on
$|\DeltaE|$, given in \Tab{tab-DeltaEcuts} that were tailored for each
individual decay mode.  For the ST analysis, we chose the candidate with
the smallest $|\DeltaE|$, if there was more than one $D$ candidate in a
particular mode.  Multiple candidates were rare in all modes except
$\Dzkpipiz$ where approximately 15\% of the events had multiple candidates.
In events with only two charged tracks that were consistent with our
requirements for $\Dzkpi$ decays, we also imposed additional requirements
to eliminate $\ep\em\to\ep\em$ and $\ep\em\to\mup\mum$ events.   Since we
are not considering any all-neutral modes in this analysis, these
requirements only affect ST yields, \ie, we would not try to find DT
candidates with the $D$ or $\Dbar$ decaying in an all neutral mode recoiling
from a
$\Dzkpi$ candidate. 

\Begtable{htb}{Requirements on $\DeltaE$ for $D$
candidates accepted for the final analysis.  The limits are set at
approximately $3\sigma$.}{tab-DeltaEcuts}
\Begtabular{lc}
Mode              &  Requirement (MeV)\\ \hline
$\Dzkpi$          &  $|\DeltaE|< 29.4$      \\       
$\Dzkpipiz$       &  $-58.3<\DeltaE<35.0$      \\       
$\Dzkpipipi$~~~   &  $|\DeltaE|<20.0$    \\       
$\Dpkpipi$        &  $|\DeltaE|<21.8$    \\       
$\Dpkspi$         &  $|\DeltaE|<26.5$    \\       
\Endtabular
\Endtable

To obtain $D\Dbar$ events for DT yields, we select only one candidate per
event per combination $D$ and $\Dbar$ decay modes.  We apply the $\DeltaE$
requirements described for the ST analyses, but do not choose the best
candidate on the basis of minimum $|\DeltaE|$.  Instead, we choose the
combination with the average of $\MD$ and $\MDbar$ -- \ie,
$\MDavg \equiv [\MD+\MDbar]/2$ -- closest to $M_D$. 
In careful studies of Monte Carlo 
events, we demonstrated that this procedure does not generate false peaks at the
$D$ mass in the $\MD$ \vs\ $\MDbar$ distributions that are narrow enough or
large enough to be confused with the DT signal.

\section{Generation and Study of Monte Carlo Events}

We used Monte Carlo simulations to develop the procedures
for measuring branching fractions and production cross sections, to understand
the response of the \cleoc\ detector, to determine parameters to use in
fits to determine data yields, and to estimate and understand possible
backgrounds.  In each case $\ep\em \to \psidprime \to D\Dbar$ events were
generated with the EvtGen program~\cite{evtgen}, and the response of
the detector to the daughters of the $D\Dbar$ decays was simulated with
GEANT~\cite{geant}.  The EvtGen program includes simulation of initial state
radiation (ISR) -- radiation of a photon by the $\ep$ and/or $\em$ before
their annihilation.  The program PHOTOS~\cite{photos} was used to simulate
final state radiation (FSR) -- radiation of photons by the charged
particles in the final state.  We simulated two types of $D$ meson decays:
\Begitem
\item  signal Monte Carlo decays, in which a $D$ or a $\Dbar$
decays in one of the 5 modes measured in this analysis, and 
\item generic Monte Carlo decays, in which a
$D$ or $\Dbar$ decays in accord with decay modes and branching fractions based
on the 2002 Particle Data Group compilation~\cite{pdg2002}. (Some tuning was
required to match inclusive distributions in data.)
\Enditem
Using these two types of simulated decays, we generated three types of Monte
Carlo events:
\Begitem
\item generic Monte Carlo events, in which both the $D$ and the
$\Dbar$ decay generically,
\item single tag signal Monte Carlo events, in which either the
$D$ or the $\Dbar$ always decays in one of the 5 modes measured in this
analysis while the $\Dbar$ or $D$, respectively, decays generically, and
\item double tag signal Monte Carlo events, in which both the $D$
and the $\Dbar$ decay in one of the 5 modes in this analysis.  
\Enditem
Our signal Monte Carlo events included simulations of ISR, but our generic Monte
Carlo events were generated before ISR was included in the simulations.  The
roles played by these Monte Carlo samples are described below.  

We applied the same selection criteria for $D$ candidates and $D\Dbar$
events in analyzing data and Monte Carlo events.   We compared many
distributions of particle kinematical quantities in data and Monte Carlo
events to asses the accuracy and reliability of the Monte Carlo simulation of
event generation and detector response.  The agreement between data and Monte
Carlo events for both charged and neutral particles was excellent for nearly
all distributions of kinematic variables that we studied.  The results of
this analysis are not sensitive to the modest discrepancies that were
observed in a few distributions.  

\section{Single Tag Efficiencies and Data Yields}

We used binned likelihood fits to ST $\MD$ distributions in Monte Carlo
events to determine experimental resolutions and ST efficiencies.  We then
used some of these experimental resolution parameters in binned likelihood
fits to data to determine ST data yields.   Four probability distribution
functions were used in  fits to ST data and ST Monte Carlo events:
\Begitem
\item Fits to mass ($m = \MD$) distributions in ST Monte Carlo events without
ISR used a relatively narrow core Gaussian $g(m;\mu,\sigma)$ to account for
beam energy spread and a small contribution from detector resolution. The mean
$\mu$, standard deviation
$\sigma$, and area
$A$ of the core Gaussian were among the parameters determined in the fits.
\item In fits to Monte Carlo events with ISR and to data, the core Gaussian
function was replaced by an inverted Crystal Ball~\cite{cbf} function 
$c(m;\mu,\sigma,\alpha,n)$ to account for the radiative tail on the high
mass side of the peak due to ISR.  This function is
identical to $g(m;\mu,\sigma)$ for $m < \mu+\alpha\sigma$, but for larger
values of $m$ the Gaussian is replaced with a radiative tail function that
depends on $\alpha$ and $n$. The parameters $\alpha$ and $n$ are determined
by fits to the mass distributions. 
\item All fits included a wider bifurcated Gaussian,
$b(m;\mu,\sigmaL,\sigmaR)$ with different standard deviations $\sigmaL$ and
$\sigmaR$ to the left and right of the peak at $m = \mu$, respectively.  In all
fits, we constrained the parameter $\mu$ in this function to be equal to the mean
of the core Gaussian.  This term models misreconstruction of
charged particle tracks, neutral pions, and neutral kaons.   The ratios
$\rL=\sigmaL/\sigma$, $\rR=\sigmaR/\sigma$ of the left and right standard
deviations of the bifurcated Gaussian to the standard deviation of the core
Gaussian and the ratio ($\rA$) of the area of the bifurcated Gaussian to the total
signal area (\ie, area of the bifurcated Gaussian plus the area of either the core
Gaussian or the Crystal Ball function) were parameters in the fits.  In fits
involving ISR, the parameters
$\mu$ and
$\sigma$ used here refer to the Gaussian parameters of the Crystal Ball function.  
\item Although combinatorial backgrounds in all $\MD$ distributions were very
small, an ARGUS background function~\cite{argusf} 
included in each of the fits to account for this contribution. The ARGUS
function $a(m;\Ez,\xi)$ is a function of the beam energy $\Ez$ and a
parameter $\xi$ determined in the fit.
\Enditem
We always fit the $\MD$ and $\MDbar$ distributions for ST events in each
signal mode individually, with the parameters of the signal distributions
constrained to be equal, but with no constraints on background sizes or
shapes. The parameters determined from $\MD$ and $\MDbar$
distributions for a given mode were in excellent agreement. The manners in
which we utilized these four contributions to fit Monte Carlo events and data
are:

\Begitem
\item Single tag yields in signal Monte Carlo events without ISR\\ 
We obtained signal Monte Carlo events without ISR from our sample of signal
Monte Carlo events with ISR by rejecting events with ISR photons with
energies $E_\gamma > 25$~keV.  We used this sample to study the shapes
of the experimental $\MD$ distributions due to detector resolution and
misreconstructions and determine the parameters of the core and bifurcated
Gaussians for use in fits to other distributions.
The means of the core Gaussians are all in excellent agreement with the masses
$\MDz$ and $\MDp$ from the PDG~\cite{PDG}.   
The widths of the core Gaussians are in the range
$\sigma = 1.25$-1.35~MeV for all modes except $\Dzkpipiz$ where 
$\sigma \approx 1.7$~MeV. The contributions of the
bifurcated Gaussians are small and their widths are somewhat larger than the
width of the core Gaussian: for all modes except $\Dzkpipiz$ we find 
$\rL = 2.0$, $\rR = 2.5$-2.6, and $\rA =3$\%-6\%; while for $\Dzkpipiz$ we find 
$\rL = 2.9$, $\rR = 3.2$, and $\rA = 11$\%.  The differences between
$\Dzkpipiz$ and the other modes are due to $\piz$ reconstruction uncertainties.  
\item Single tag yields in signal Monte Carlo events with ISR\\
We used all single tag Monte Carlo events to determine the efficiencies
$\effi$ for detecting ST decays in the signal modes and to determine the
contribution of ISR to the shape of the $\MD$ peak.  For each signal mode, we
used the Gaussian parameters, $\rL$,  $\rR$, and $\rA$ determined from the ST
events without ISR.   We used the total areas (sum of the two Gaussians and
the radiative tail) of the $\MD$ peaks to determine the ST efficiencies
$\effi$. 
\item Single tag yields in generic Monte Carlo\\
We used our large sample of generic Monte Carlo data (without
ISR) to verify the effectiveness of our fitting procedures, to study
combinatorial backgrounds, and to provide input data to test the branching
fraction fit program.  We fixed the values of the parameters $\sigma$,
$\rL$, and $\rR$ to the values obtained from signal Monte Carlo events
without ISR.
\item Single tag yields in data\\
We fit the $\MD$ distribution for a given decay mode in ST data using the
Gaussian parameters, $\rL$, $\rR$, and $\rA$ determined from the signal Monte
Carlo events without ISR.  From the fit, we determined 
$\sigma$, $A$, $\alpha$, $n$, and $\xi$.  The yields are the
sums of the contributions from the two Gaussians and the radiative tail.  The
data and the results of these fits are illustrated in
Figs.~\ref{fig-stdzdata}  and \ref{fig-stdpdata} for 
$\Dz\Dzbar$ and $\Dp\Dm$ candidate events, respectively.   The linear
plots demonstrate that the misreconstruction contributions and
the combinatorial backgrounds are small, and the semi-logarithmic plots
demonstrate that the signals and backgrounds are modeled well by the fits.  
\Enditem   

\Begtable{htb}{Single tag data yields and efficiencies and their
statistical uncertainties.}{tab-STYieldsAndEffs}
\Begtabular{lcc}
$D$ or $\Dbar$ Mode &  Yield ($10^3$) & ~~~Efficiency (\%)~~~ \\ \hline
$\Dzkpi$               &  $5.14\pm 0.07$ & $65.1\pm 0.6$ \\
$\Dzbarkpi$            &  $5.16\pm 0.08$ & $66.3\pm 0.6$ \\
$\Dzkpipiz$            &  $9.62\pm 0.12$ & $33.6\pm 0.4$ \\
$\Dzbarkpipiz$         &  $9.58\pm 0.12$ & $34.0\pm 0.4$ \\
$\Dzkpipipi$           &  $7.39\pm 0.10$ & $45.1\pm 0.5$ \\
$\Dzbarkpipipi$ ~~~~~~ &  $7.39\pm 0.10$ & $45.5\pm 0.5$ \\
$\Dpkpipi$             &  $7.58\pm 0.09$ & $52.2\pm 0.5$ \\
$\Dmkpipi$             &  $7.57\pm 0.09$ & $51.9\pm 0.5$ \\
$\Dpkspi$              &  $1.09\pm 0.04$ & $45.6\pm 0.5$ \\
$\Dmkspi$              &  $1.12\pm 0.04$ & $45.9\pm 0.5$ \\
\Endtabular
\Endtable

\Tab{tab-STYieldsAndEffs} gives the ST yields in data and the ST
efficiencies calculated from the signal Monte Carlo events with ISR.  These
quantities are used in the fit, described in \Sec{sec-bffit}, for branching
fractions and numbers of
$D\Dbar$ events.  

\section{Double Tag Efficiencies and Data Yields}\label{sec-dtyields}

The components that must be included in the DT fits are illustrated in
\Fig{fig-scatter}, which shows distributions of $\MDbar$ \vs\ $\MD$ for DT event
candidates that are reconstructed in the $\Dzbarkpipiz$ and $\Dzkpipiz$ modes. 
Figure~\ref{fig-scatter}(a) shows all candidates (including multiple candidates)
after imposition of the $\DeltaE$ requirement described in
\Sec{sec-data&cuts}.  The principal features of this two-dimensional
distribution are:
\Begitem
\item
There is an obvious strong signal concentration with a complicated shape
in the region surrounding $\MDbar = \MD = \MDz$.
\item There is a radiative tail above the signal peak at $45^\circ$ to the
axes.  This correlation is due to the fact that -- neglecting measurement and
reconstruction errors -- the values of $\MD$ and $\MDbar$ calculated using
the beam energy will both to be too large by the same amount if energy was
lost due to ISR.
\item There are faint horizontal and vertical bands at $\MDbar = \MDz$ and
$\MD = \MDz$, respectively.  The events in these bands are events in which
the $\Dzbar$ ($\Dz$) candidate was reconstructed correctly, but the 
$\Dz$ ($\Dzbar$) was misreconstructed.  
\item There is a somewhat more visible band below the peak at $45^\circ$ to
the axes that continues through the signal region and the radiative tail.   
This band is populated by two sources of background:
\Begitem
\item  There are $D\Dbar$ events, in which all of the particles were found and
reconstructed reasonably accurately, but one or more particles from the real
$\Dz$ were interchanged with the corresponding particles from the
$\Dzbar$, \eg, the two $\piz$s were interchanged. 
\item There are also continuum events, \ie, annihilations into $u\bar{u}$,
$d\bar{d}$, and $s\bar{s}$ quark pairs.
\Enditem
\Enditem
As illustrated in
\Fig{fig-scatter}(b), all three bands below the peak are depopulated
noticeably after choosing the candidate with the average mass $\MDavg$
nearest to $\MDz$.  Since careful studies of this choice using Monte Carlo
simulations of generic $D\Dbar$ decays and continuum events showed no evidence
of
false peaks in the region of the $D$ mass, the net effect of this procedure is
reduction of background without introduction of a false signal.

We included five terms in unbinned likelihood fits to account for these
features of the two-dimensional $\MDbar$ \vs\ $\MD$ distributions.  These
terms were functions of $m = \MD$, $\mbar = \MDbar$, 
$\mavg = \MDavg = [\MD + \MDbar]/2$, and $\deltam = [\MD - \MDbar]/2$.  The
terms are:
\Begitem
\item $c(\mavg;\mu,\sigma,\alpha,n)\,g(\deltam;\mu,\sigmadelta)$, a Crystal
Ball function of $\mavg$ multiplied by a Gaussian function of $\deltam$.  The
Crystal Ball function models nearly all of the signal, including the
radiative tail, while the Gaussian function models the
effect of measurement errors in the direction transverse to $\mavg$.  
In these fits, $\mu$, $\sigma$, and $\sigmadelta$ were allowed to float. 
The parameters $\mu$ and $\sigma$ also appear in other terms; they float in
all terms, but their values are constrained to be identical in all terms. 
The parameters $\alpha$ and $n$ used in DT fits to data or Monte Carlo events
with ISR, were fixed to the averages over all modes of the parameters
obtained in ST fits to data or Monte Carlo events with ISR, respectively. 
In fits to Monte Carlo events without ISR the Crystal Ball function is
replaced with a pure Gaussian.  The values of Gaussian and Crystal Ball
parameters obtained in DT yields are close to -- but not exactly equal to -- 
those obtained in the ST fits.
\item
$b(m;\mu,\rL\sigma,\rR\sigma)\,b(\mbar;\mu,\rL\sigma,\rR\sigma)$, the
product of two bifurcated Gaussian functions.  These functions account for
misreconstructed signal events as they did individually in ST fits.  The
ratios $\rL$ and $\rR$ were obtained from the fits to ST Monte Carlo
events without ISR.  The signal contribution is the sum of
this contribution and the Crystal Ball contribution.
\item $a(\mavg;\Ez,\xi)\,g(\deltam;\mu,\sigmac)$, the product of an ARGUS
background function of $\mavg$ and a Gaussian function of $\deltam$.  This
term models continuum background and any residual mispartitioned events
remaining after choosing the value $\MDavg$ closest to $\MD$ in events with
multiple candidates.  The Gaussian parameter $\sigmac$ is determined from
continuum Monte Carlo events, and $\xi$ is allowed to float in
the fit.
\item $a(m;\Ez,\xim)\,g(\mbar;\mu,\sigmab)$, the product of an ARGUS
function of $m$ and a Gaussian function of $\mbar$.  This term models
the horizontal band in the scatter plot.  The values of $\xim$ and $\sigmab$
are the identical in this term and the next.  The value of $\sigmab$ is
determined from DT Monte Carlo events without ISR and is the same for all
modes, while $\xim$ is allowed to float in the fits.
\item $a(\mbar;\Ez,\xim)\,g(m;\mu,\sigmab)$, the product of an ARGUS
function of $m$ and a Gaussian function of $\mbar$.  This term models
the vertical band in the scatter plot.  The value of $\xim$ in this term is
constrained to be equal to the value of $\xim$ in the term for the
horizontal band.
\Enditem 

\Begtable{htb}{Double tag data yields and efficiencies and their
statistical uncertainties.}{tab-DTYieldsAndEffs}
\Begtabular{llcc}
$D$ Mode    & $\Dbar$ Mode    & Yield ($10^2$)   & ~~~Efficiency (\%)~~~ \\ \hline
$\Dzkpi$          & $\Dzbarkpi$         & $1.09\pm 0.11$   & $42.6\pm 0.5$ \\
$\Dzkpipiz$       & $\Dzbarkpipiz$      & $4.84\pm 0.23$   & $12.1\pm 0.3$ \\
$\Dzkpipipi$      & $\Dzbarkpipipi$     & $2.80\pm 0.17$   & $20.8\pm 0.4$ \\
$\Dzkpi$          & $\Dzbarkpipiz$      & $2.45\pm 0.16$   & $23.2\pm 0.4$ \\
$\Dzkpipiz$       & $\Dzbarkpi$         & $2.62\pm 0.16$   & $22.6\pm 0.4$ \\
$\Dzkpi$          & $\Dzbarkpipipi$     & $2.05\pm 0.14$   & $29.6\pm 0.4$ \\
$\Dzkpipipi$      & $\Dzbarkpi$         & $1.97\pm 0.14$   & $29.6\pm 0.4$ \\
$\Dzkpipiz$       & $\Dzbarkpipipi$     & $3.59\pm 0.20$   & $15.2\pm 0.3$ \\
$\Dzkpipipi$ ~~~~ & $\Dzbarkpipiz$ ~~~~ & $3.40\pm 0.19$   & $15.5\pm 0.3$ \\
$\Dpkpipi$        & $\Dmkpipi$          & $3.79\pm 0.20$   & $26.7\pm 0.4$ \\
$\Dpkspi$         & $\Dmkspi$           & $0.090\pm 0.030$ & $20.6\pm 0.4$ \\
$\Dpkpipi$        & $\Dmkspi$           & $0.609\pm 0.079$ & $23.7\pm 0.4$ \\
$\Dpkspi$         & $\Dmkpipi$          & $0.530\pm 0.073$ & $23.9\pm 0.4$ \\
\Endtabular
\Endtable

\noindent Figures~\ref{fig-dtdzdata} and \ref{fig-dtdpdata} show the projection of
each of the ``diagonal''  $\MDbar$ \vs\ $\MD$ distributions projected onto the
$\MD$ axes.  The fit in each figure is the sum of all terms in the fit projected
on the
$\MD$ axis.  
\Tab{tab-DTYieldsAndEffs} gives the DT yields in data and the corresponding
DT efficiencies calculated from the signal Monte Carlo events with ISR.  These
quantities are used in the fit -- described in \Sec{sec-bffit} for branching
fractions and numbers of $D\Dbar$ events.  

\section{Efficiencies in Data and Monte Carlo Events}

We estimated efficiencies for reconstructing $D$ decays using
Monte Carlo simulations.  For precision measurements, we must also understand
the accuracy with which the Monte Carlo events simulate
tracking efficiencies.   
Using  partial reconstruction techniques, we measure tracking efficiencies --
in both data and Monte Carlo events -- for charged pions and kaons, as well as
the efficiency for detecting
$\KS$ and $\piz$ mesons.  For
example, we fully reconstruct one $\Dz$ and then combine it with one or
more of the other particles in the event to partially reconstruct the
$\Dzbar$, leaving out one particle -- $\pipm$, $\Kpm$, or 
$\KS$ -- for which we wish to measure the efficiency.  We calculate the
square of the missing mass ($\Mmsq$) of the combination, which should
peak at the  mass squared of the missing particle.  We then check whether the
missing particle is actually detected or not, and we plot the missing mass
squared for both classes of events.  Fitting the signal peaks tells us what
fraction of particles were actually detected.  We also use similar
techniques with 
$\psiprime\to\Jpsi\,\pip\pim$ and $\psiprime\to\Jpsi\,\piz\piz$ events to
measure $\pipm$ tracking efficiency at lower momenta and $\piz$ efficiency,
respectively.  

We illustrate this technique for measuring $\Kpm$ tracking efficiencies
in \Fig{fig-keff}, where -- for both data and Monte Carlo events --
we reconstruct $\Dz$ mesons in the modes, $\Dzkpi$, $\Dzkpipiz$, and
$\Dzkpipipi$.   We use the requirements $|M(\Dz) - \MDz| < 5$~\Mevcsq\ and
$|\DeltaE| < 25$~MeV to obtain a clean sample of $\Dz$ decays. 
Figures~\ref{fig-keff}(a) and \ref{fig-keff}(b) show the $\Mmsq$ distributions
for  data and Monte Carlo events, respectively, when the $\Kpm$ was
found, while Figs.~\ref{fig-keff}(c) and \ref{fig-keff}(d) show the
corresponding $\Mmsq$ distributions for events in which the $\Kpm$ was
not found.  When the $\Kpm$ is found, there is a clean peak with
little background in both data and Monte Carlo events.  Each of these
distributions is fit with two Gaussians of the same mean but different
widths, and a small constant
background.  The widths of the Gaussians are somewhat different in
data and Monte Carlo events.  The $\Mmsq$ distributions for events in
which the $\Kpm$ was not found are more complex.  There is a peak at
$M_{\pip}^2$ due to $\Dz \to \pip\pim$ decays, and a peak at
$M_{\Kp}^2$ due to $\Dzbarkpi$ decays.  The shoulder on the right of
each figure is due to $\Dzbarkpipiz$ decays in which only the $\pim$ is
detected; its shape is well approximated by an error function. 
Finally, there is a linearly rising shape which describes the
remaining backgrounds.  We determine the 
parameters for the shapes of these backgrounds
in fits to signal Monte Carlo events for these processes. 
We fit the data in Figs.~\ref{fig-keff}(c) and \ref{fig-keff}(d) using these
background shape parameters and the widths of the Gaussian parameters
determined in fits to events in which the $\Kpm$ was detected.    

For $\pipm$ and $\Kpm$, we find evidence for slightly higher tracking
efficiencies in data than in Monte Carlo events.  
Based on these preliminary studies, we increase the efficiencies determined
from the Monte Carlo simulations by a factor of $1.03 \pm 0.03$ for {\it each}
charged track candidate used in the event.  In our preliminary studies
of $\piz$ efficiencies, we see no evidence
for an efficiency difference between data and Monte Carlo events.
Also, we find no evidence for a correction for $\KS$ decays beyond the factor
of $1.03^2$ for the two charged daughter pions in the decay. 

\section{Systematic Errors}

\Begtable{htb}{Systematic uncertainties and the quantities to which they are
applied in the branching fraction fit.  Uncorrelated, additive uncertainties
are given in the first section, and correlated, multiplicative uncertainties
in the second.}{tab-systematics}
\Begtabular{lcc}
Source & Fractional Uncertainty (\%) & Quantity \\ \hline
Data processing           & 0.3      & All yields \\
Yield fit functions       & 0.1--2.9 & All yields \\
Background bias           & 2.5      & DT yields \\
Double DCSD interference  & 0.8      & Neutral DT yields \\ \hline
Detector simulation       & 3.0      & Tracking efficiencies \\
                          & 3.0      & $\KS$ efficiencies \\
                          & 4.4      & $\piz$ efficiencies \\
                          & 0.3      & $\pi^\pm$ PID efficiencies \\
                          & 1.0      & $K^\pm$ PID efficiencies \\
Trigger simulation        & 0.3      & ST efficiencies \\
Final state radiation     & 0.5      & $D$ efficiencies \\
$|\Delta E|$ requirement  & 1.0      & $D$ efficiencies, correlated by decay \\
Resonant substructure     & 3.0      & $\Dzkpipipi$ efficiencies \\
\Endtabular
\Endtable

We take systematic uncertainties into account directly in the
branching fraction fit. Table~\ref{tab-systematics} lists the
uncertainties that we included and a brief description of each contribution
follows.
\Begitem
\item Data processing\\
Potential losses of events in data processing has been limited to 0.3\% and is
assessed by reanalyzing the data with slightly different software
configurations.
\item Yield fit functions\\
We gauge the sensitivity of the ST and DT yields to variations in the $M(D)$
fit functions by repeating the fits with different values of $\rA$,
$\rR$, and $\rL$ as well as with different functional forms (such as a
symmetric
Gaussian instead of a bifurcated Gaussian for the wide signal component) and
comparing efficiency-corrected yields.
\item Background bias\\
We estimate possible biases in the background parametrization from the
agreement
of the fit functions with the data. We find the discrepancies to be negligible
for ST yields and as large as 2.5\% for DT yields.  Additional bias in the DT
yields could have been introduced by the procedure for selecting
the best candidate per event in double tags.  However, as described in 
\Sec{sec-dtyields}, we searched for this type of background in simulated
continuum
and generic $D\Dbar$ Monte Carlo events and found no evidence of such a
contamination.
\item Double DCSD interference\\
In the neutral DT modes, the Cabibbo-allowed amplitudes can
interfere with amplitudes where both $\Dz$ and $\Dzbar$ undergo
doubly Cabibbo-suppressed decays (DCSD).  We include an uncertainty to account
for the unknown phase of this interference.

\item Detector simulation -- tracking, $\KS$, and $\piz$ efficiencies\\
We estimate uncertainties due to differences between efficiencies in data and
those estimated in Monte Carlo simulations using the partial reconstruction
technique described in the previous section.

\item Detector simulation -- particle identification (PID) efficiencies\\
Particle identification efficiencies are studied by reconstructing decays
with unambiguous particle content, such as $\Dz\to K^0_S\pi^+\pi^-$ and
$\phi\to K^+ K^-$.  We also use $\Dzkpipiz$, where the $K^-$ and $\pi^+$ are
distinguished kinematically.  The efficiencies in data are well-simulated by
the Monte Carlo, and we assign correlated uncertainties of 0.3\% and 1.0\% to
each $\pi^+$ and $K^+$, respectively. We do not assign these uncertainties 
to $K^0_S$ daughters, because they are not subjected to the $\pi^+$
identification requirements.

\newpage

\item Trigger simulation\\
We assign a conservative uncertainty in the trigger simulation efficiency
taken from the trigger inefficiency we find in simulations of the
$\Dzkpipiz$ mode.

\item Final state radiation\\
In Monte Carlo simulations, final state radiation typically reduces DT
efficiencies by a factor of approximately two times the reduction of ST
efficiencies. This leads to branching fraction values larger by 0.5\% to 2\%
than they would be without including FSR in the Monte Carlo simulations. 
We have verified the accuracy of the FSR simulation
to roughly 10\% of itself using
$\psi(2S)\to J/\psi\pi\pi$ decays, where
$J/\psi\to\mu^+\mu^-\gamma$.
For these preliminary results, we assign conservative uncertainties of 
0.5\% to ST efficiencies and 1.0\% to DT efficiencies, correlated across all
modes.  This results in a $\pm 0.5$\% correlated systematic error in branching
fractions.

\item $|\DeltaE|$ requirement\\
Discrepancies in detector resolution between data and Monte Carlo simulations
can result in differences between the efficiencies of the $\DeltaE$ requirement
in data and Monte Carlo events. No evidence for such discrepancies has been
found, and we include a systematic uncertainty of 1.0\% for ST modes and 2.0\%
for DT modes, correlated by decay process.

\item Resonant substructure\\
The resonant substructure of $\Dzkpipipi$ in data is found
to disagree with that in Monte Carlo simulations, introducing a possible
fractional bias in efficiency of 3.0\%.
We apply a correlated systematic uncertainty of this size to all ST and DT
modes with a $\Dzkpipipi$ decay.

\Enditem

For the $e^+e^-\to D\bar D$ cross section measurements, we include additional
uncertainties from the luminosity measurement (3.0\%) and from variations in
the center-of-mass energy.  We estimate the latter uncertainty by calculating
$\NDzDzbar$ and $\NDpDm$ in data taken at different values of
$\Ecm$ in a 4 MeV region on the $\psi(3770)$.  We find fractional
variations of 2.5\% in $\NDzDzbar$ and 1.4\% in $\NDpDm$, and include
these numbers as systematic uncertainties.

Finally, we note that -- although there are many contributions to the systematic
errors listed in \Tab{tab-systematics} -- the contribution of tracking
efficiencies dominates in this preliminary result.  This is due to our
conservative assignment of an error of $\pm 0.03n$, where $n$ is the number of
tracks in a ST or DT mode for this uncertainty.

\section{Results of the Branching Fraction Fits}\label{sec-bffit}

To determine the branching fractions $\BDzkpi$, $\BDzkpipiz$,
$\BDzkpipipi$, $\BDpkpipi$, and $\BDpkspi$ as well as $\NDzDzbar$ and
$\NDpDm$, we measure event yields and efficiencies for the ten ST modes
and thirteen DT
modes, given in Tables~\ref{tab-STYieldsAndEffs} and~\ref{tab-DTYieldsAndEffs}.
In the branching fraction fit, we correct these event yields not only for
efficiency but also for crossfeed among the ST and DT modes and backgrounds
from other $D$ decays.  The estimated crossfeed and background contributions
induce
yield adjustments of 2\% at most.  Their dependence on the fit parameters is
taken into account both in the yield subtraction and in the $\chi^2$
minimization.
In addition to the correlated and uncorrelated systematic uncertainties, the
statistical uncertainties on the yields, efficiencies, and background
branching fractions are also included in the fit.

We validated the algorithm and the performance of the branching fit -- as well as
our entire analysis procedure -- by measuring the branching fractions in generic
Monte Carlo events.  Since the generic Monte Carlo simulations were done without
ISR, we used efficiencies obtained from signal Monte Carlo events without ISR. 
We find that the results of this procedure were in excellent agreement with the
branching fractions used by EvtGen in generating the events.  Furthermore, the
generic Monte Carlo sample has an order of magnitude more events than our data,
so the statistical errors in this test were about a factor of three smaller than
the errors in the branching fractions obtained from data.  The systematic errors
were also substantially smaller than those estimated for data, so the agreement
between measured and generated branching fractions of the generic Monte Carlo
events is a very stringent test of our entire analysis procedure. 

\Begtable{htb}{Preliminary fitted branching fractions and $D\Dbar$ pair
yields.  Uncertainties are statistical and systematic,
respectively.}{tab-dataResults}
\Begtabular{lccc}
Parameter & Fitted Value & Fractional & Fractional  \\[-0.6ex]
&& ~~~Stat. Error~~~ & Syst. Error \\ \hline
$\NDzDzbar$               & $(1.98\pm 0.04\pm 0.03)\times 10^5$ & 1.9\% & 1.5\% \\
${\cal B}(\Dzkpi)$        & $0.0392\pm 0.0008\pm 0.0023$        & 2.1\% & 5.8\% \\
${\cal B}(\Dzkpipiz)$     & $0.143\pm 0.003\pm 0.010$           & 2.0\% & 7.2\% \\
${\cal B}(\Dzkpipipi)$    & $0.081\pm 0.002\pm 0.009$           & 2.1\% &11.6\% \\
$\NDpDm$                  & $(1.48\pm 0.06\pm 0.04)\times 10^5$ & 4.3\% & 2.4\% \\
${\cal B}(\Dpkpipi)$      & $0.098\pm 0.004\pm 0.008$           & 4.4\% & 8.5\% \\
${\cal B}(\Dpkspi)$       & $0.0161\pm 0.0008\pm 0.0015$        & 4.9\% & 9.1\% \\
\hline
${{\cal B}(\Dzkpipiz)}/{{\cal B}(\Dzkpi)}$
                          & $3.64\pm 0.05\pm 0.17$              & 1.3\% & 4.8\% \\
${{\cal B}(\Dzkpipipi)}/{{\cal B}(\Dzkpi)}$ ~
                          & $2.05\pm 0.03\pm 0.14$              & 1.4\% & 6.8\% \\
${{\cal B}(\Dpkspi)}/{{\cal B}(\Dpkpipi)}$
                          & $0.164\pm 0.004\pm 0.006$           & 2.5\% & 3.7\% \\
\Endtabular
\Endtable

The results of the data fit are shown in Table~\ref{tab-dataResults}.
The $\chi^2$ of the fit is
8.9 for 16 degrees of freedom, corresponding to a confidence level of 92\%.
To obtain the separate contributions from statistical and systematic
uncertainties, we repeat the fit without any systematic inputs and take the
quadrature difference of uncertainties.
All five branching fractions are consistent with -- but also higher than --
the current PDG averages~\cite{PDG}.  If no FSR is included in the simulations to
calculate signal efficiencies, then all the branching fractions would be 0.5\% to
2\% lower.
In addition, we compute ratios of branching fractions to the reference branching
fractions, which have higher precision than  the constituent
branching fractions, and these also agree with the PDG averages.  These ratios are
also sensitive to final state radiation, and -- without these corrections -- all
three ratios would be 1\% to 2\% higher.

\Begtable{htb}{Correlation matrix, including systematic uncertainties,
for the branching fractions determined from the fit.}{tab-dataCorrelationMatrix}
\Begtabular{l|ccccccc}
& $\NDzDzbar$ & $K^-\pi^+$ & $K^-\pi^+\pi^0$ & $K^-\pi^+\pi^-\pi^+$ &
	$\NDpDm$ & $K^-\pi^+\pi^+$ & $K^0_S\pi^+$ \\
\hline
$\NDzDzbar$ & 1 & $-0.37$ & $-0.26$ & $-0.21$
	& $0.02$ & $-0.02$ & $-0.02$ \\
$K^-\pi^+$  & & 1 & 0.75 &  0.90 & 0.00 & 0.76 & 0.69 \\
$K^-\pi^+\pi^0$ & & & 1 & 0.73 & 0.00 & 0.62 & 0.57 \\
$K^-\pi^+\pi^-\pi^+$ & & & & 1 & 0.00 & 0.80 & 0.73 \\ 
$\NDpDm$ & & & & & 1 & $-0.49$ & $-0.47$ \\
$K^-\pi^+\pi^+$ & & & & & & 1 & 0.90 \\
$K^0_S\pi^+$ & & & & & & & 1\\
\Endtabular
\Endtable

The correlation matrix for the seven fit parameters is given in
Table~\ref{tab-dataCorrelationMatrix}.  In the absence of
systematic uncertainties, there would be no correlation between the charged
and neutral $D$ parameters.  That these correlations are, in fact, large
indicates that the precision of our branching fraction measurements is
limited by our understanding of systematic effects.

We obtain the $e^+e^-\to D\bar D$ cross sections by dividing the fitted values
of $\NDzDzbar$ and $\NDpDm$ by the luminosity collected on the
$\psi(3770)$, which we determine to be $\Lum = 57.2 \pm 1.7$~\pbinv.
Thus, at
$\Ecm=3773$ MeV, we find preliminary values of the production cross sections,
\begin{eqnarray*}
\sigma( e^+e^-\to D^0\bar D^0 ) &=& (3.47\pm 0.07\pm 0.15) \ {\rm nb} \\
\sigma( e^+e^-\to D^+ D^- ) &=& (2.59\pm 0.11\pm 0.11) \ {\rm nb} \\
\sigma( e^+e^-\to D\bar D ) &=& (6.06\pm 0.13\pm 0.22) \ {\rm nb} \\
\sigma( e^+e^-\to D^+ D^- ) / \sigma( e^+e^-\to D^0\bar D^0 ) &=&
	\hspace*{0.4em} 0.75\pm 0.04\pm 0.02,
\end{eqnarray*}
where the uncertainties are statistical and systematic, respectively.
The charged and neutral cross sections have a correlation coefficient of 0.32
stemming from the common use of the luminosity measurement.

\section{Conclusions}

We report preliminary measurements of three $\Dz$ and two $\Dp$ branching
fractions in a sample of nearly 60\pbinv\ of $\ep\em\to D\Dbar$ data obtained at
$\Ecm = 3.77$~GeV.  The results are presented in \Tab{tab-dataResults} and the
correlation coefficients among the results are given in
\Tab{tab-dataCorrelationMatrix}.  
We find branching fractions in agreement with -- but somewhat higher -- than
those in the PDG~\cite{PDG} summary.
We note that the Monte Carlo simulations used in calculating efficiencies in this
analysis included final state radiation.  However, the branching fractions used
in the Particle Data Group averages do not include this effect.  If we had not
included final state radiation in our simulations, branching fractions would have
been 0.5\% to 2\% lower.  At this preliminary stage in our analysis,
the systematic errors are factors of approximately 2 to 5 larger than the
statistical errors.   Our statistical errors for branching fractions and ratios
of branching fractions are -- in nearly all cases -- less than the statistical
errors in the individual measurements that contributed to the PDG averages.  

Our techniques for event selection, for fitting data and Monte Carlo
distributions to obtain yields and efficiencies, and for fitting the results to
obtain branching fractions are validated with generic Monte Carlo events to a
much higher precision than the statistical errors from the current data sample. 
This indicates that we can improve the results from these data substantially with
more detailed understanding of any possible systematic differences between our
data and our Monte Carlo simulations.
Our systematic errors are dominated by the uncertainty of $0.03n$, where $n$
is the number of charged tracks in a ST or DT mode, that we assign for the
difference between tracking efficiency in data and Monte Carlo simulations.    
Many of the systematic uncertainties, such as those for tracking 
and particle identification efficiencies, will be improved with larger
data samples.

Our preliminary measurements of the production cross sections
$\sigma(\Dz\Dzbar)$, $\sigma(\Dp\Dm)$, and $\sigma(D\Dbar)$ are in good agreement
with the cross sections measured by the MARK~III collaboration~\cite{markiii-2}
and in excellent agreement with recent BES results reported at the 2004
Moriond conference~\cite{bes-sigmaddbar}.

\section{Acknowledgements}

We gratefully acknowledge the effort of the CESR staff 
in providing us with
excellent luminosity and running conditions.
This work was supported by 
the National Science Foundation,
the U.S. Department of Energy,
the Research Corporation,
and the Texas Advanced Research Program.

\Begfigure{hb}
 \includegraphics[width=0.45\textwidth]{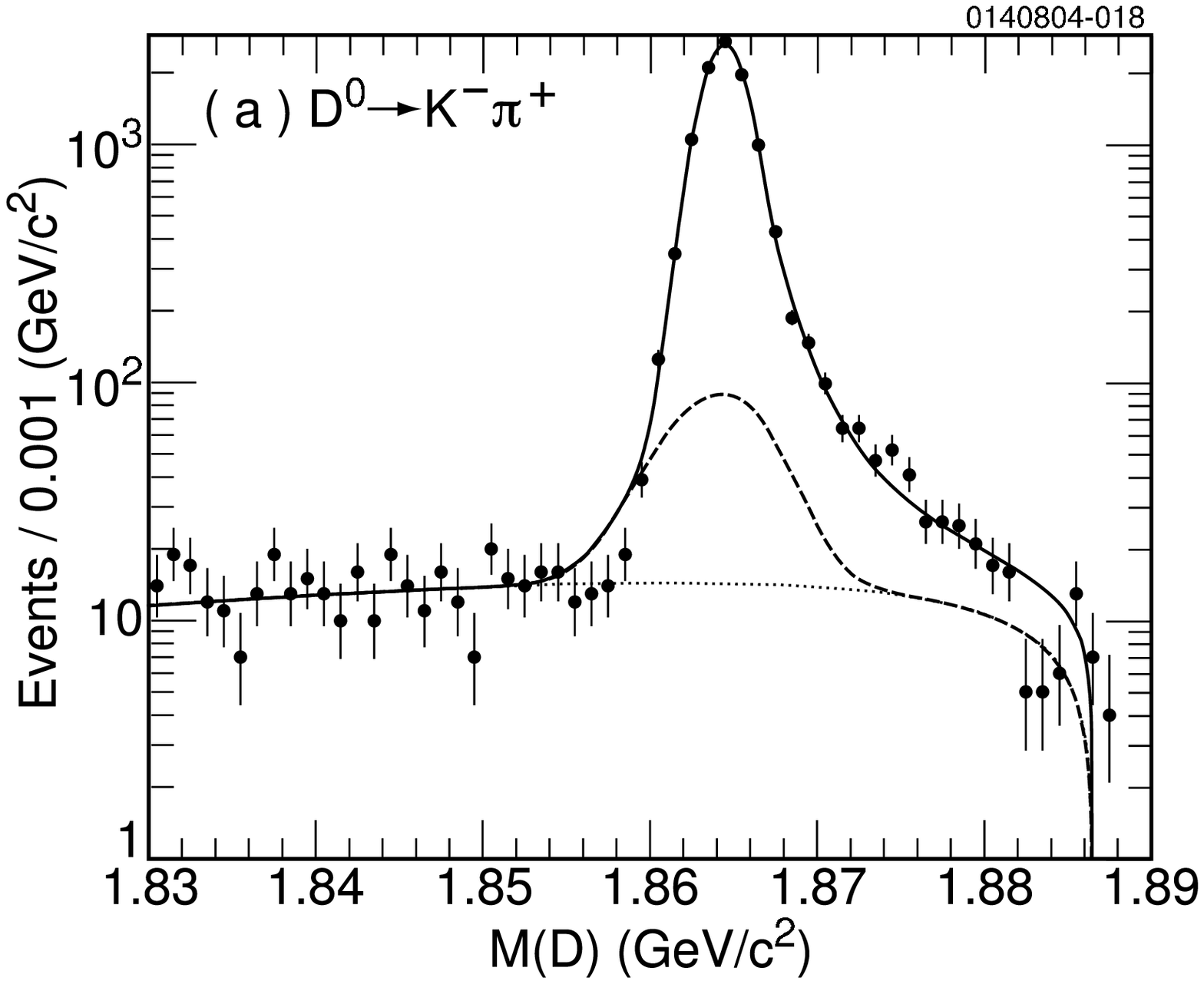}
 \includegraphics[width=0.45\textwidth]{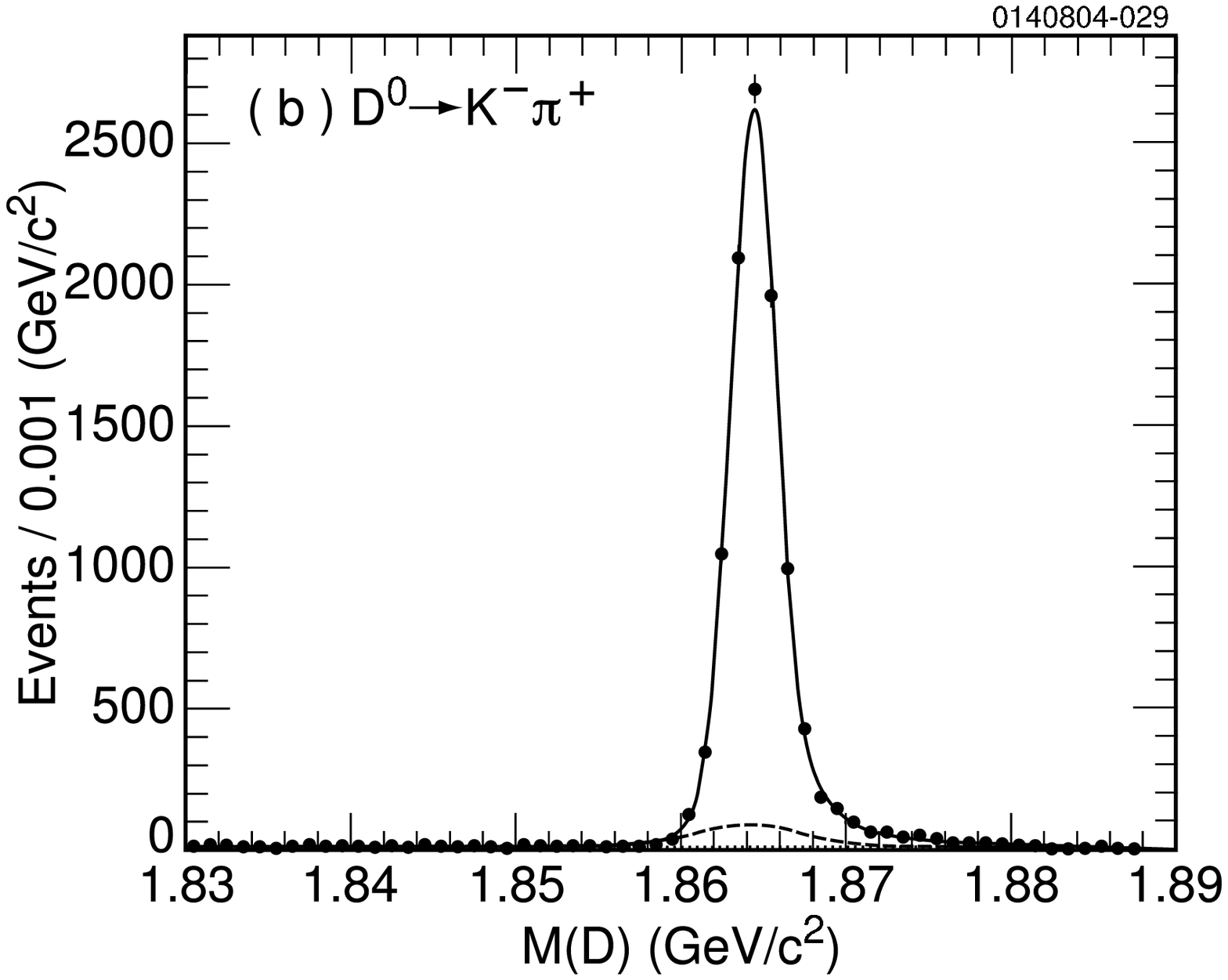}\\[3ex]
 \includegraphics[width=0.45\textwidth]{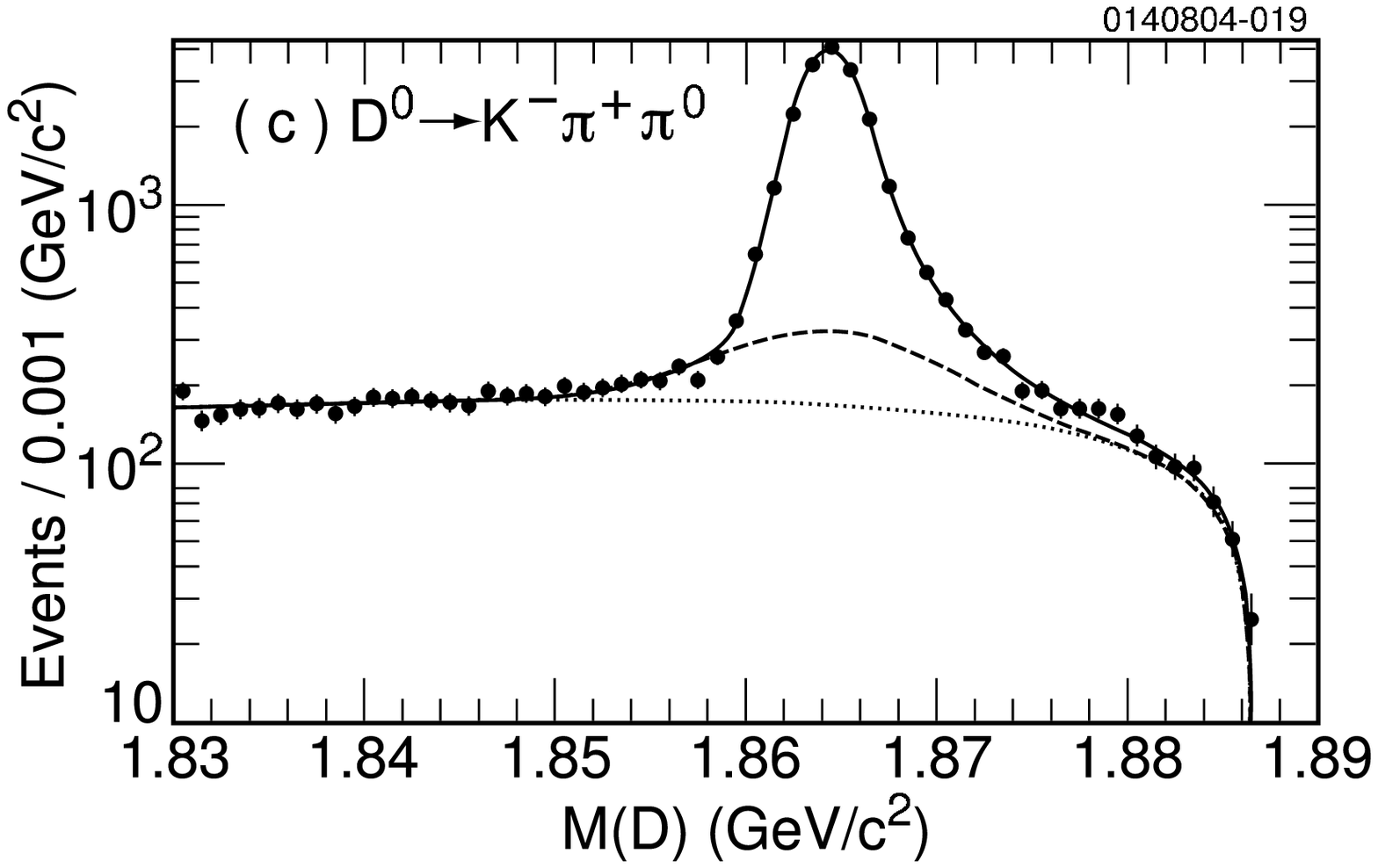}
 \includegraphics[width=0.45\textwidth]{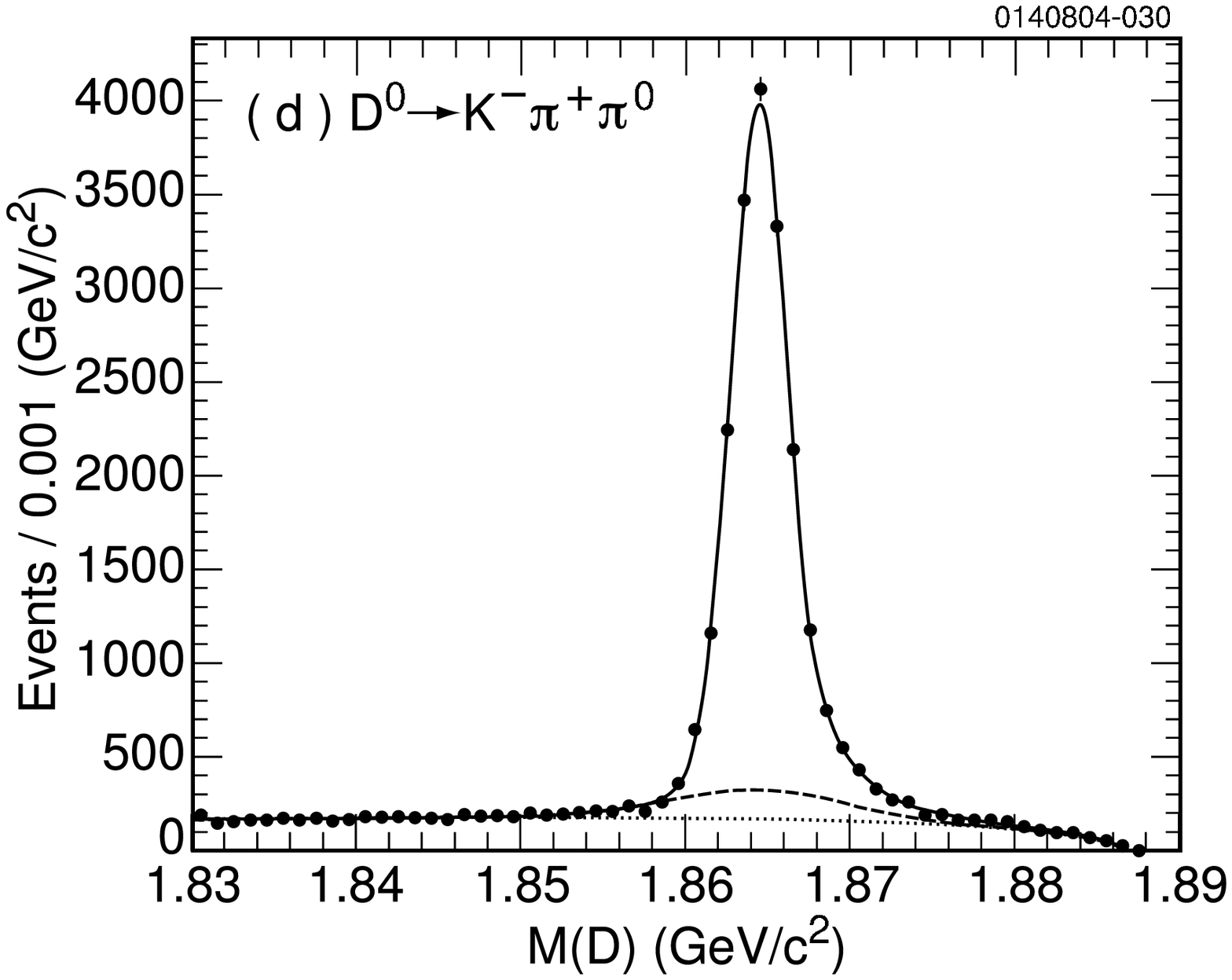}\\[3ex]
 \includegraphics[width=0.45\textwidth]{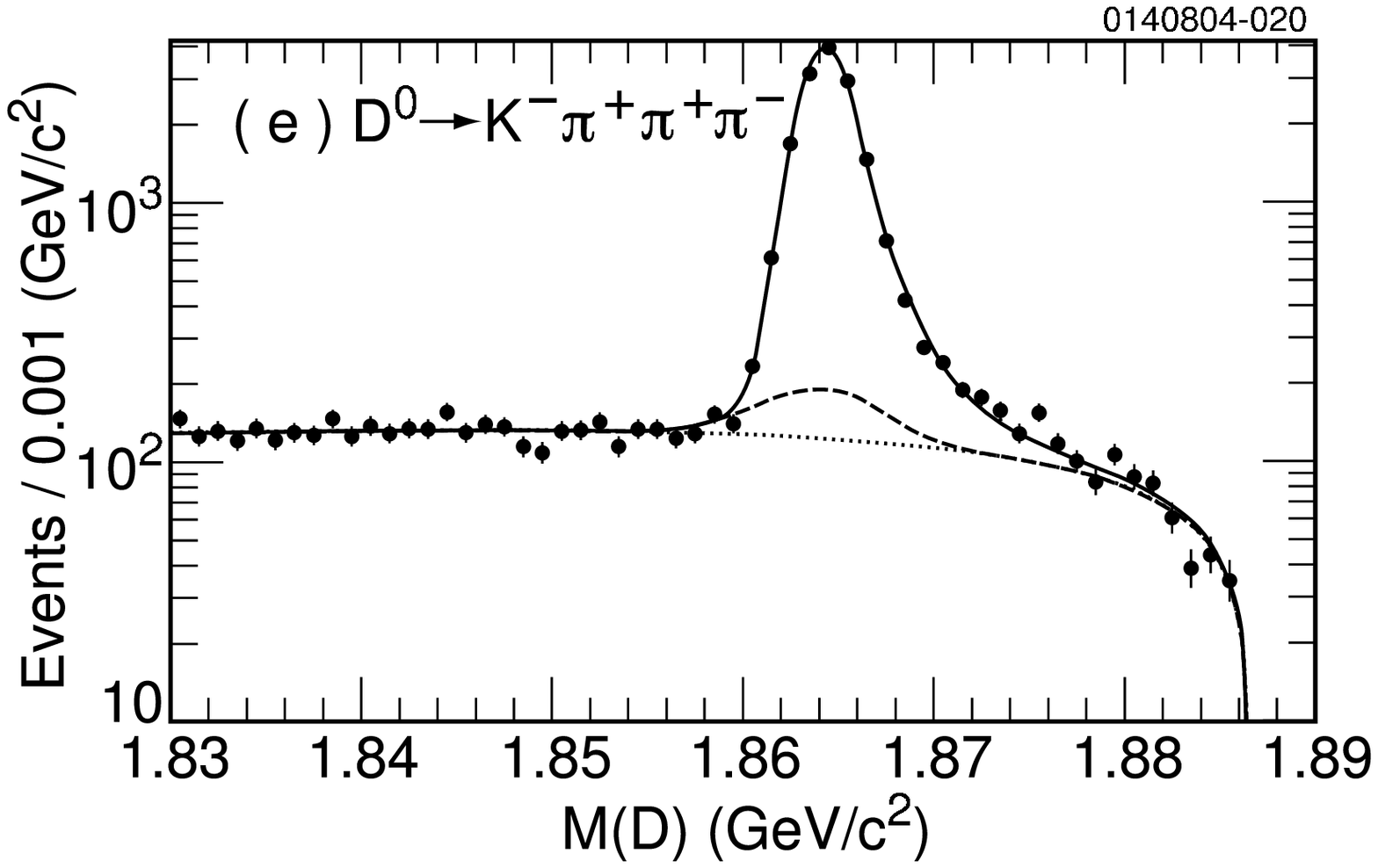}
 \includegraphics[width=0.45\textwidth]{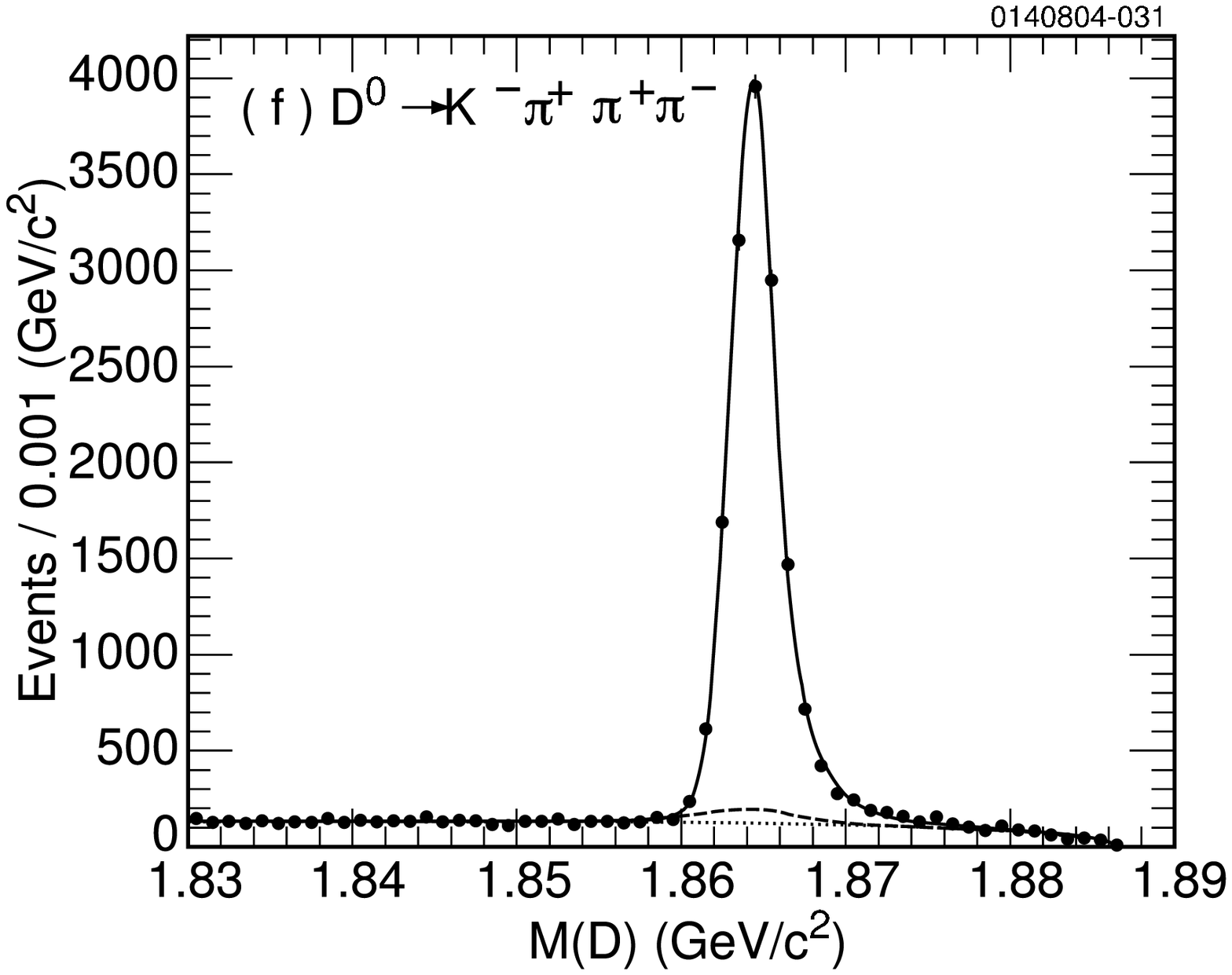}
\Endfigure{Distributions of calculated $\MD$ values for single tag neutral
$D$ candidates in data with $\Dz$ and $\Dzbar$ combined in each mode.  The
points are data and the curves are fits to the data.  In each plot, the
dotted curve shows the contribution of the ARGUS background function, the
dashed curve shows the  contribution of the bifurcated Gaussian, and the
solid curve shows the total, including the Crystal Ball function.  (a) and
(b) illustrate $\Dzkpi$ candidates, (c) and (d) illustrate $\Dzkpipiz$
candidates, and (e) and (f) illustrate $\Dzkpipipi$ candidates.}
{fig-stdzdata}

\clearpage       
 
\Begfigure{tb}
 \includegraphics[width=0.45\textwidth]{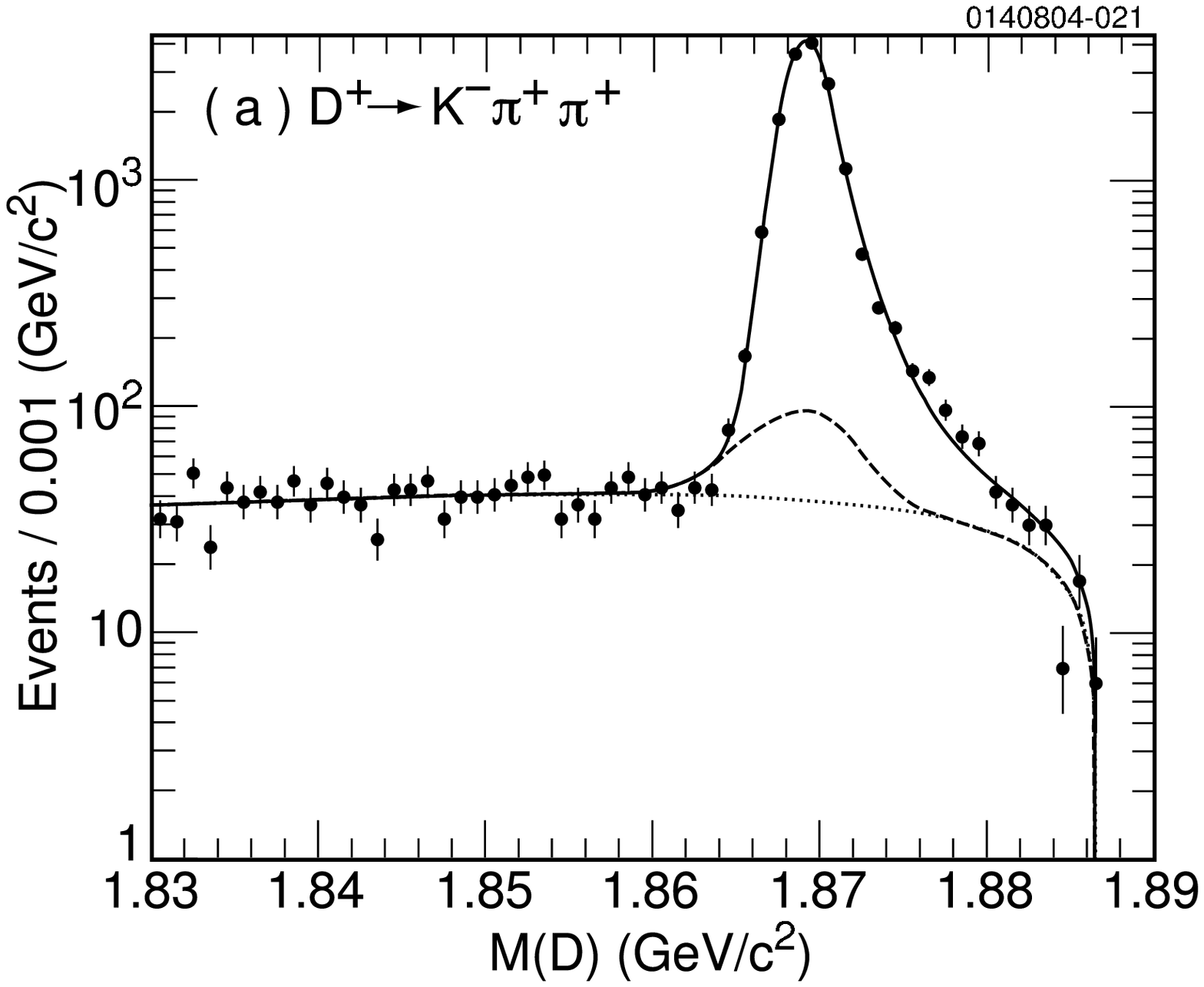}
 \includegraphics[width=0.45\textwidth]{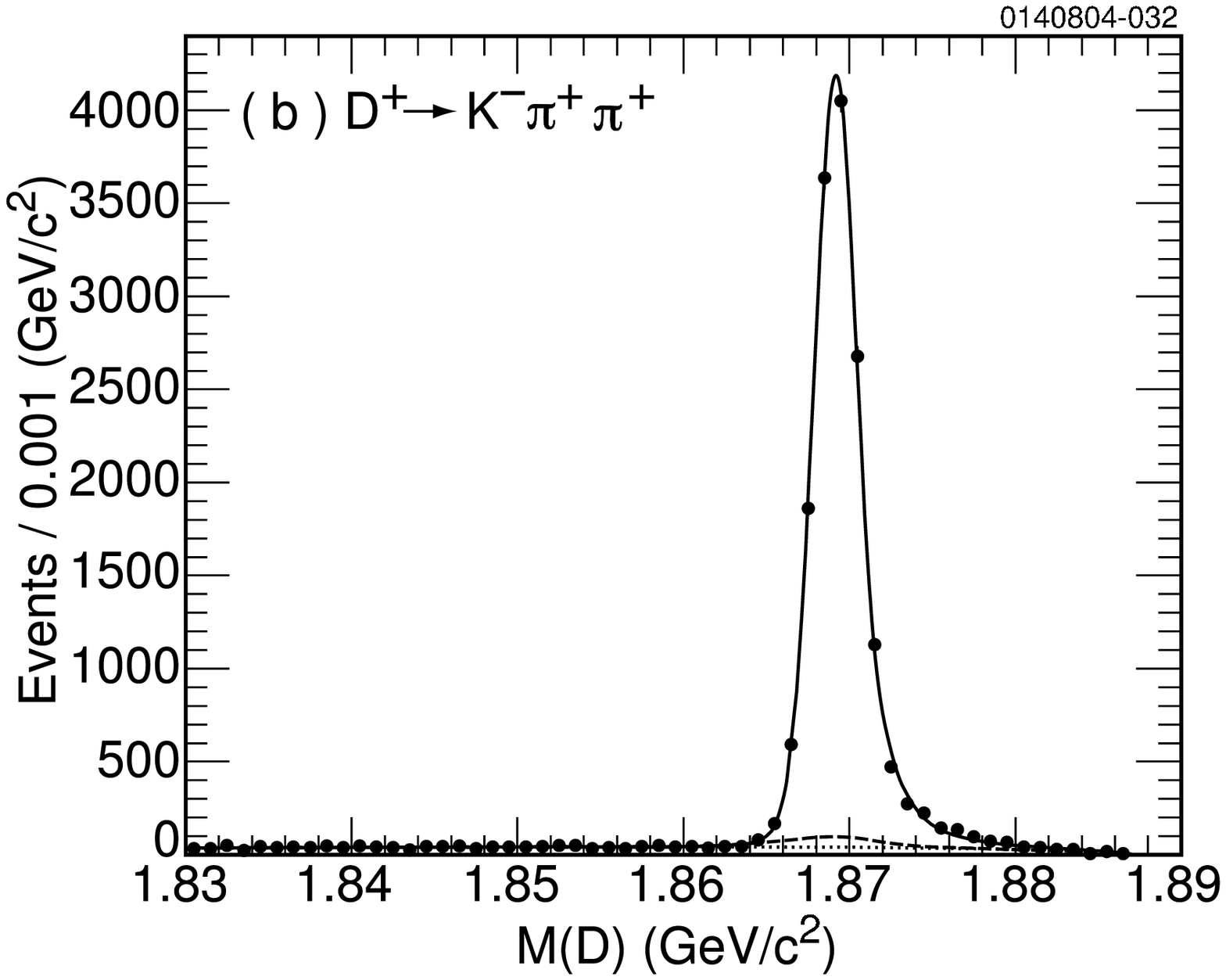}\\[3ex]
 \includegraphics[width=0.45\textwidth]{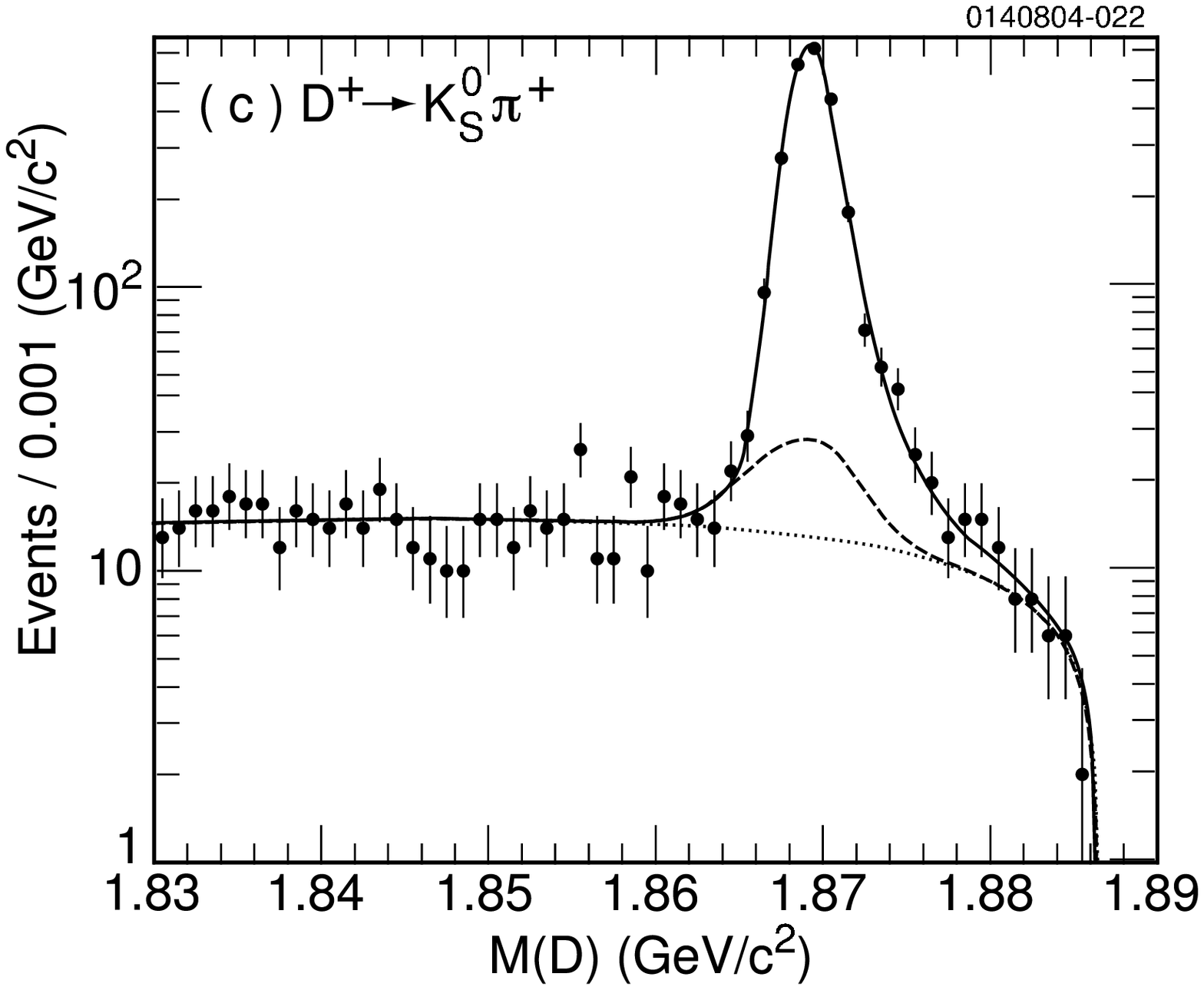}
 \includegraphics[width=0.45\textwidth]{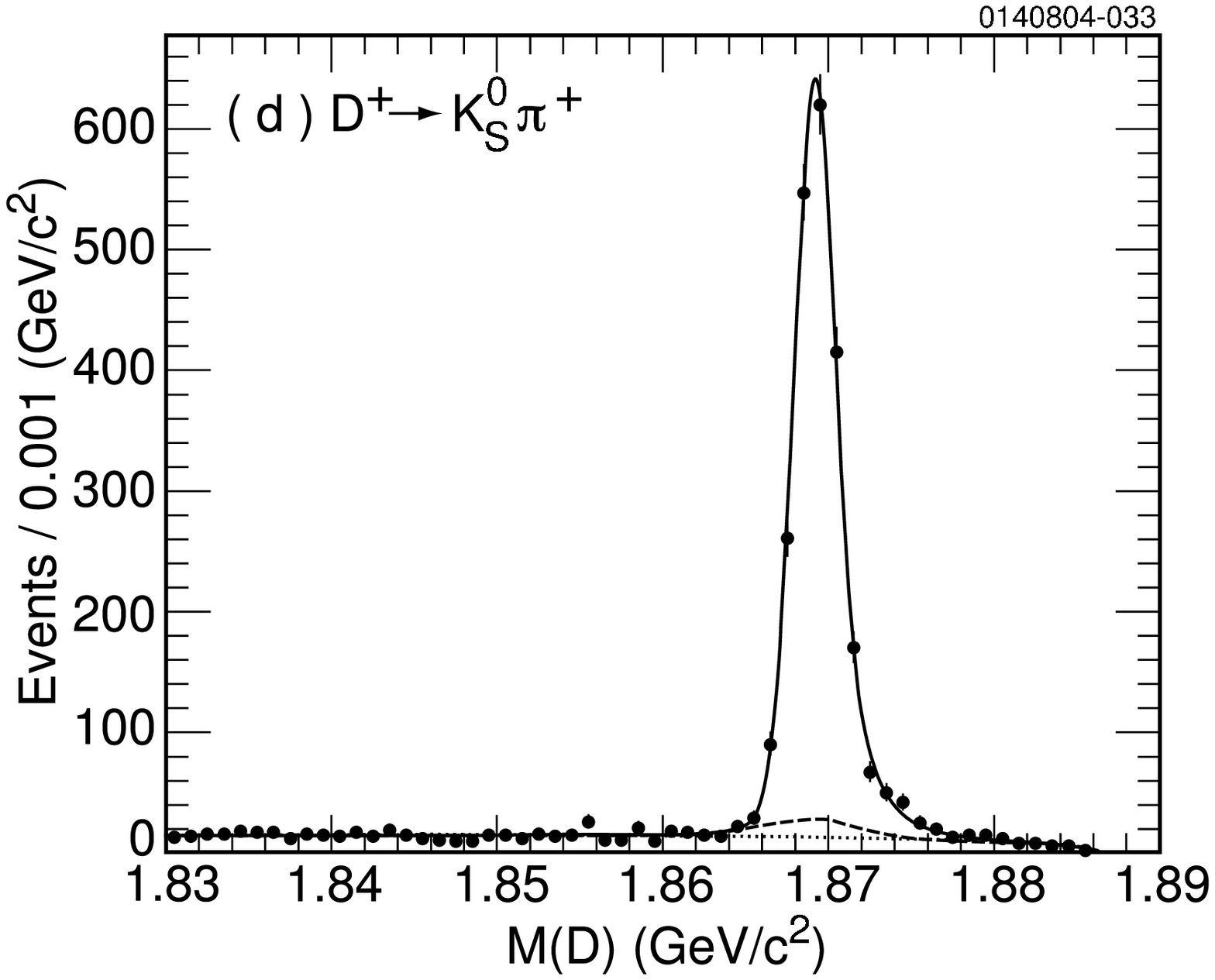}
\Endfigure{Distributions of calculated $\MD$ values for single tag charged
$D$ candidates in data with $\Dp$ and $\Dm$ combined in each mode.  The
points are data and the curves are fits to the data.  In each plot, the
dotted curve shows the contribution of the ARGUS background function, the
dashed curve shows the  contribution of the bifurcated Gaussian, and the
solid curve shows the total, including the Crystal Ball function.    (a) and
(b) illustrate $\Dpkpipi$ candidates and (c) and (d) illustrate $\Dpkspi$
candidates.}{fig-stdpdata}       
 
\clearpage       
 
\Begfigure{tb}
 \includegraphics[width=0.96\textwidth]{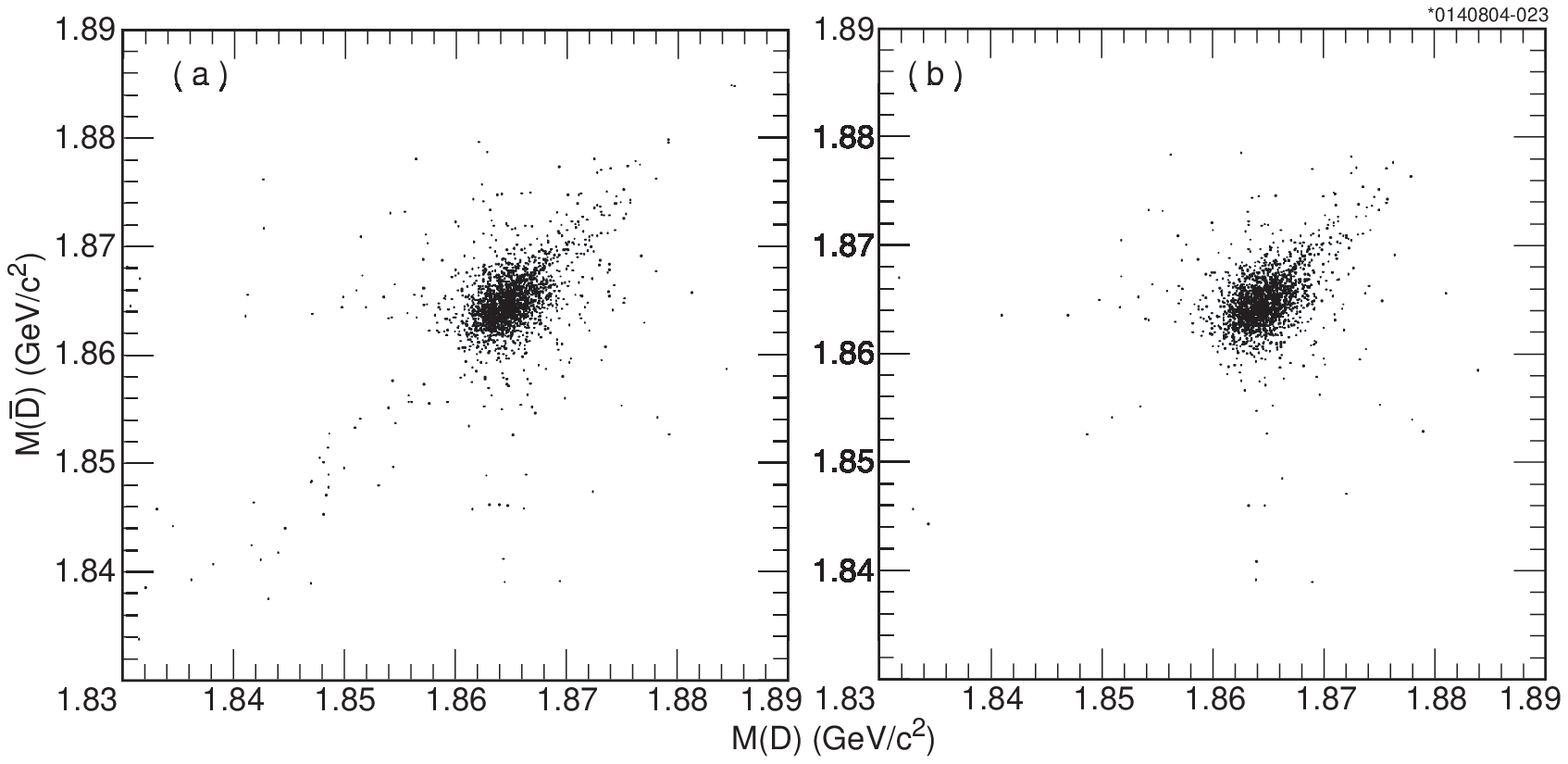}
\Endfigure{
Scatter plot of $\MDbar$ \vs\ $\MD$ values for $\Dzbarkpipiz$ \vs\
$\Dzkpipiz$ DT candidates from simulated events. (a) illustrates all
candidates after  imposing the $\DeltaE$
requirement for each candidate given in \Tab{tab-DeltaEcuts}.  (b)
illustrates the candidates remaining after choosing the candidate with the
average $\MDavg$ closest to
$\MDz$ in events with multiple candidates.} {fig-scatter}

\clearpage       
 
\Begfigure{tb}
 \includegraphics[width=0.45\textwidth]{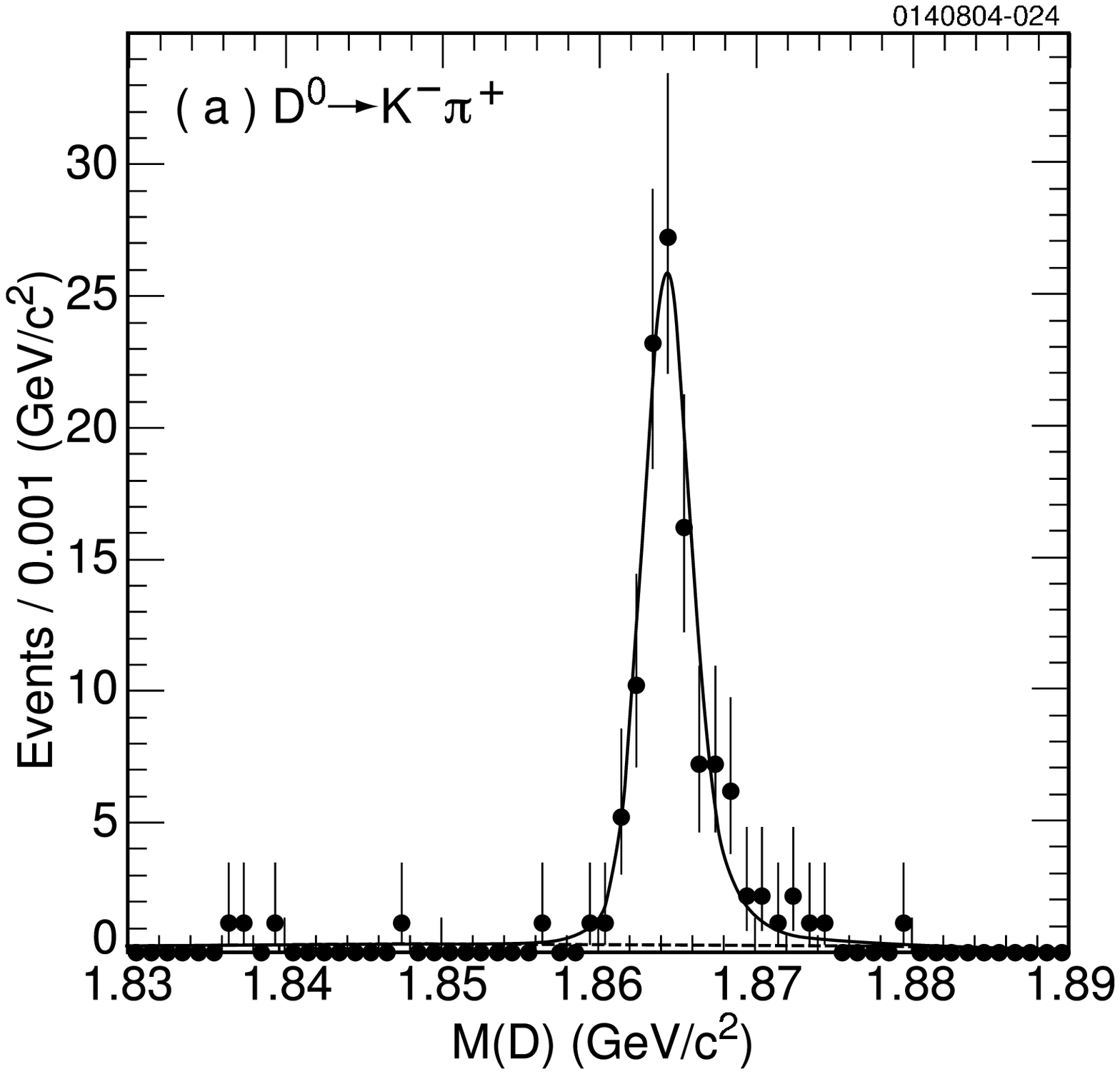}
 \includegraphics[width=0.45\textwidth]{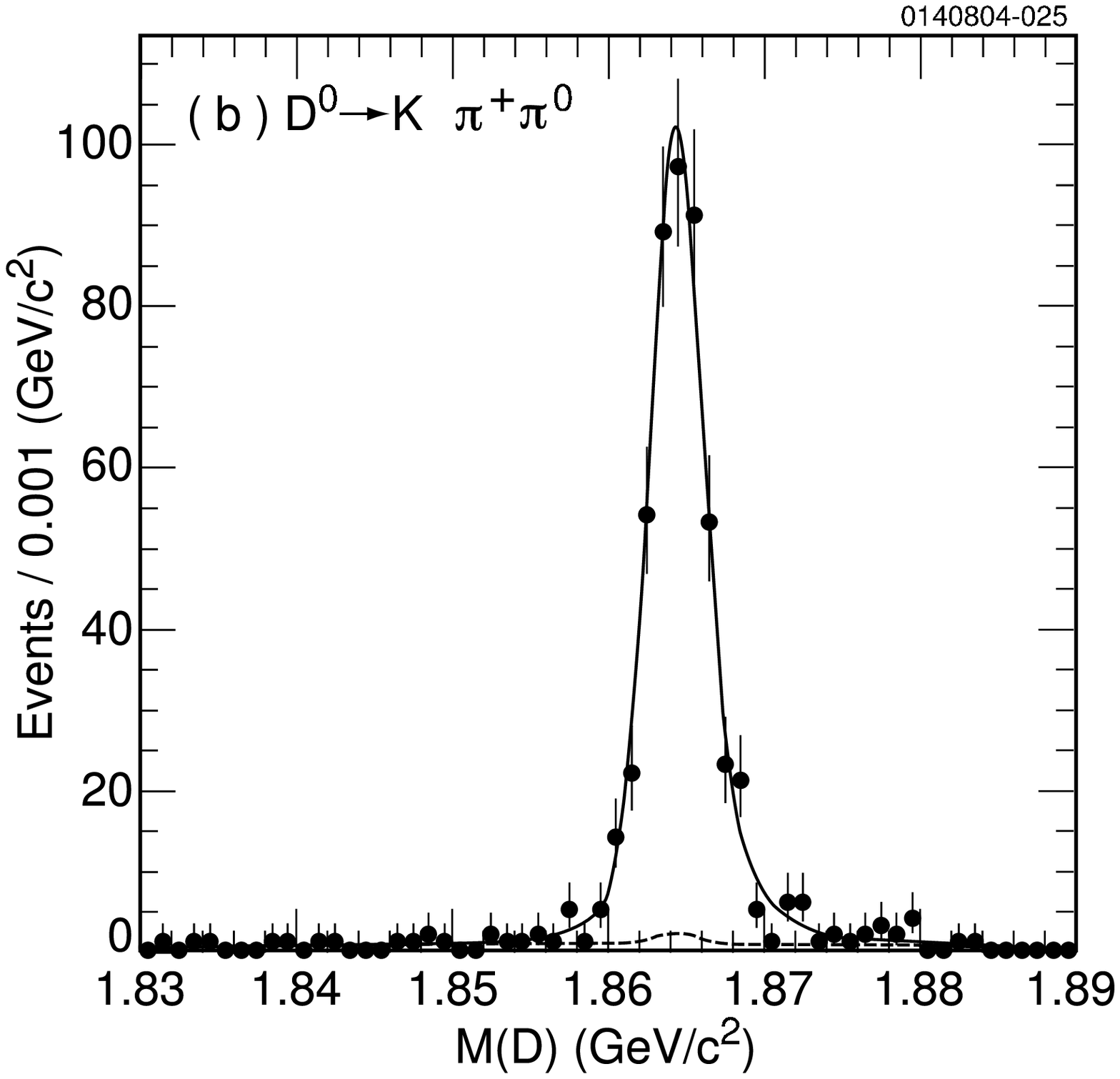}\\[3ex]
 \includegraphics[width=0.45\textwidth]{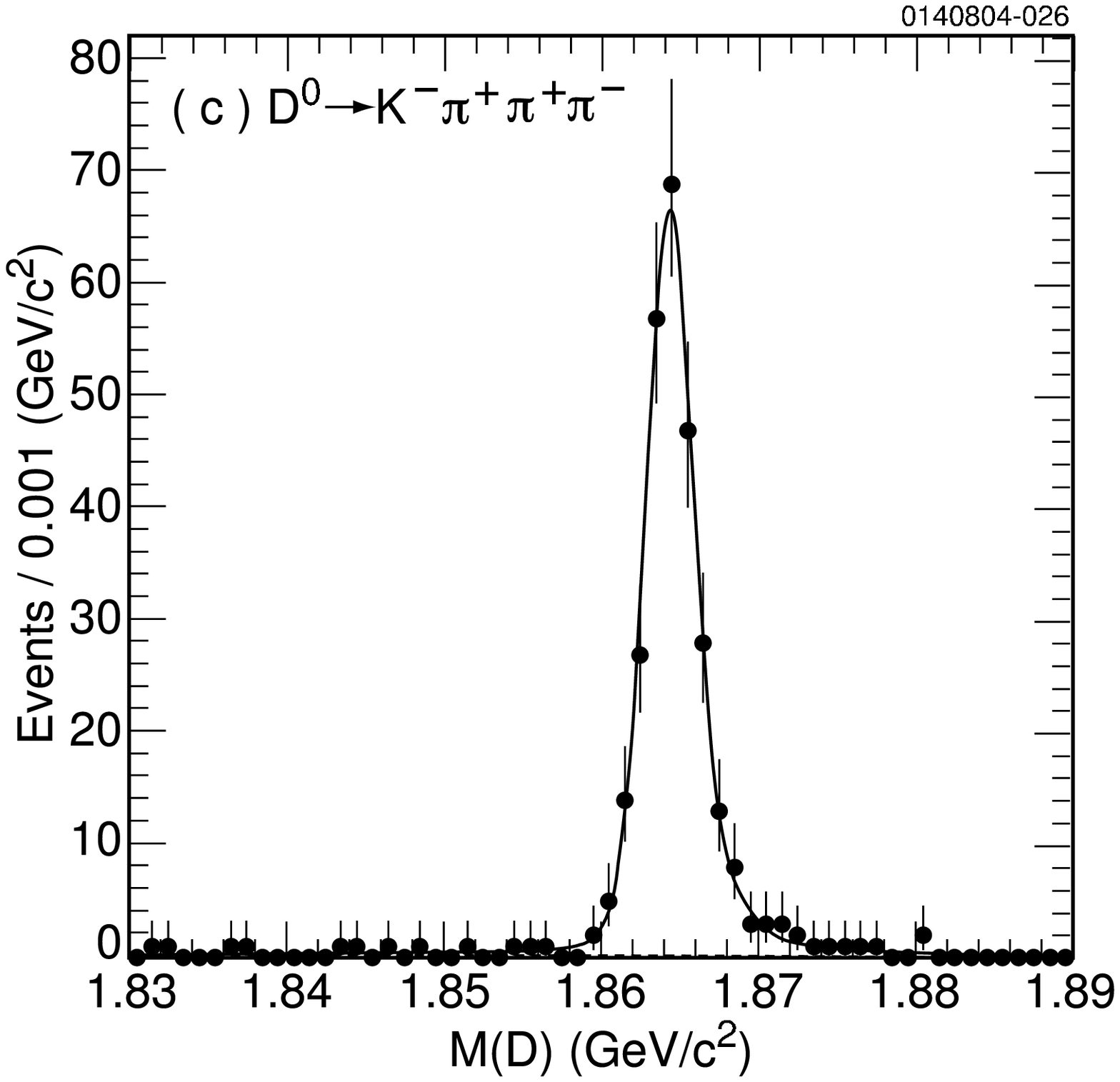}
\Endfigure{
Projections of double tag $\Dz$-$\Dzbar$ candidate masses on the $\MD$ axis,
with the $\Dz$ and $\Dzbar$ reconstructed in corresponding charge conjugate
modes.  Projections of the candidate masses on the $\MDbar$ axis are nearly
identical.  (a) illustrates $\Dzkpi$ candidates, (b) illustrates $\Dzkpipiz$
candidates, and (c) illustrates $\Dzkpipipi$ candidates.} {fig-dtdzdata}

\Begfigure{tb}
 \includegraphics[width=0.45\textwidth]{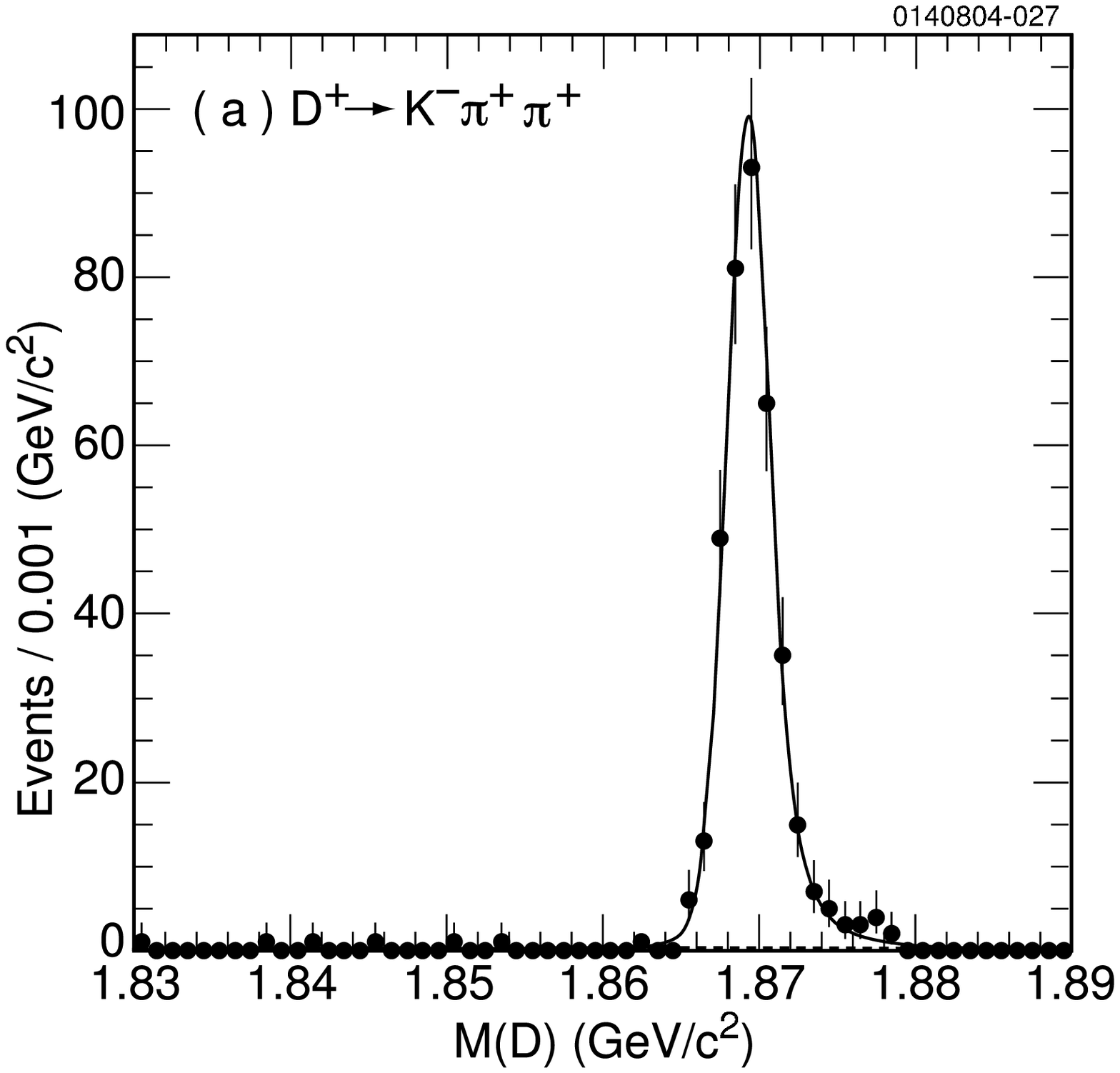}
 \includegraphics[width=0.45\textwidth]{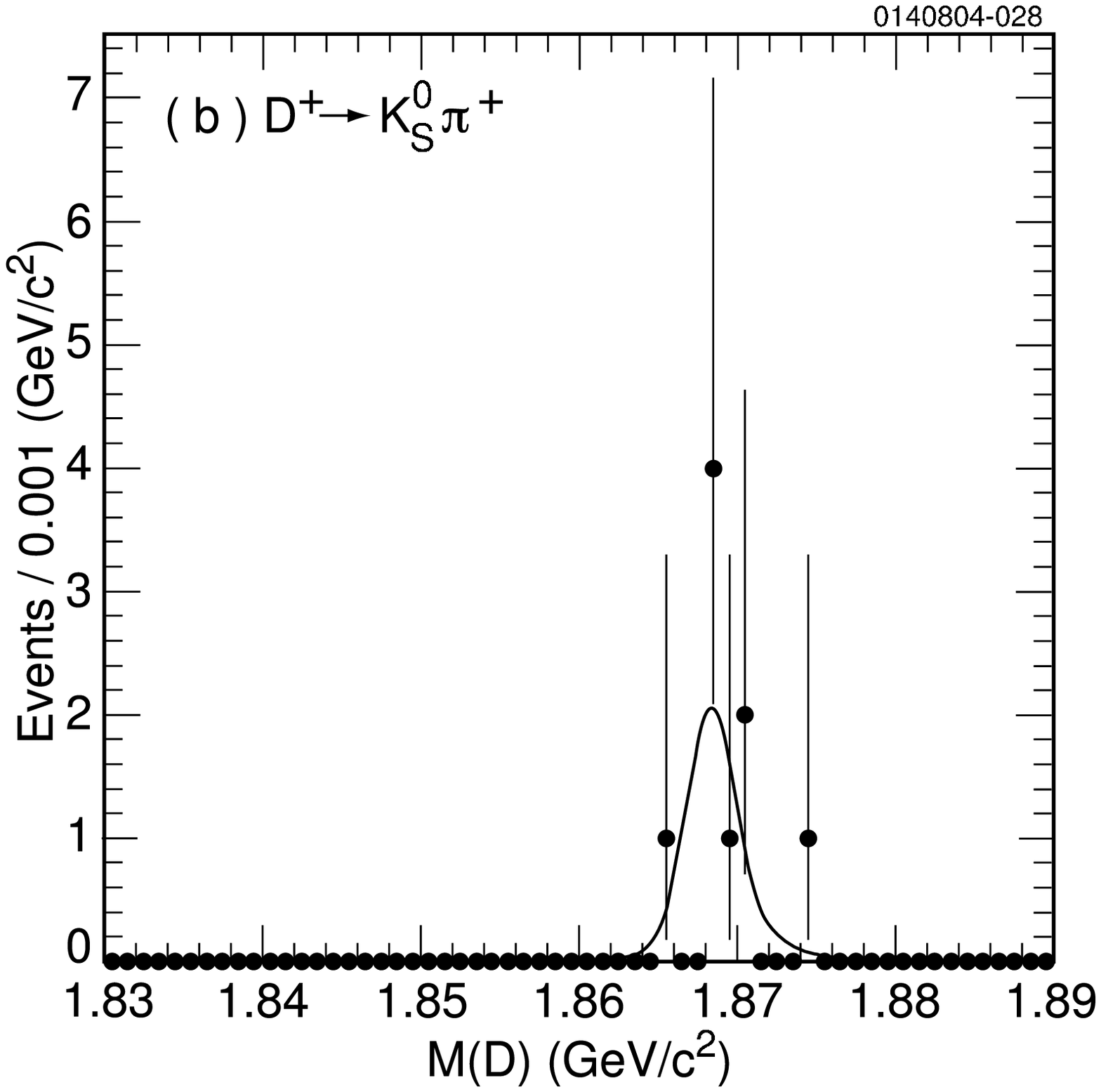}
\Endfigure{
Projections of double tag $\Dp$-$\Dm$ candidate masses on the $\MD$ axis,
with the $\Dp$ and $\Dm$ reconstructed in corresponding charge conjugate modes. 
Projections of the candidate masses on the $\MDbar$ axis are nearly
identical.  (a) illustrates $\Dpkpipi$ candidates and (b) illustrates
$\Dpkspi$ candidates.} {fig-dtdpdata}

\clearpage       
 
\Begfigure{htb}
\includegraphics[width=0.48\textwidth]{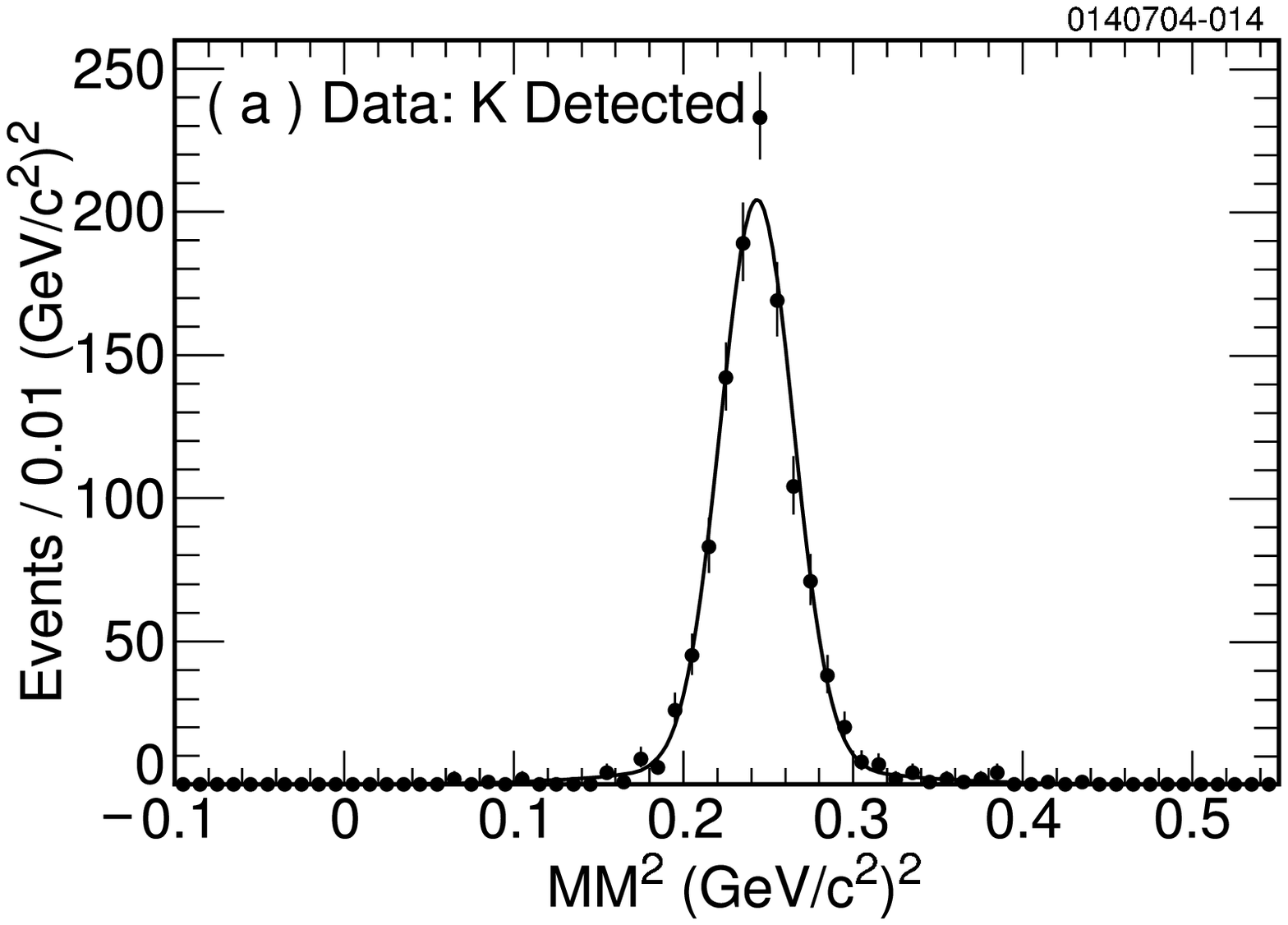}
\includegraphics[width=0.48\textwidth]{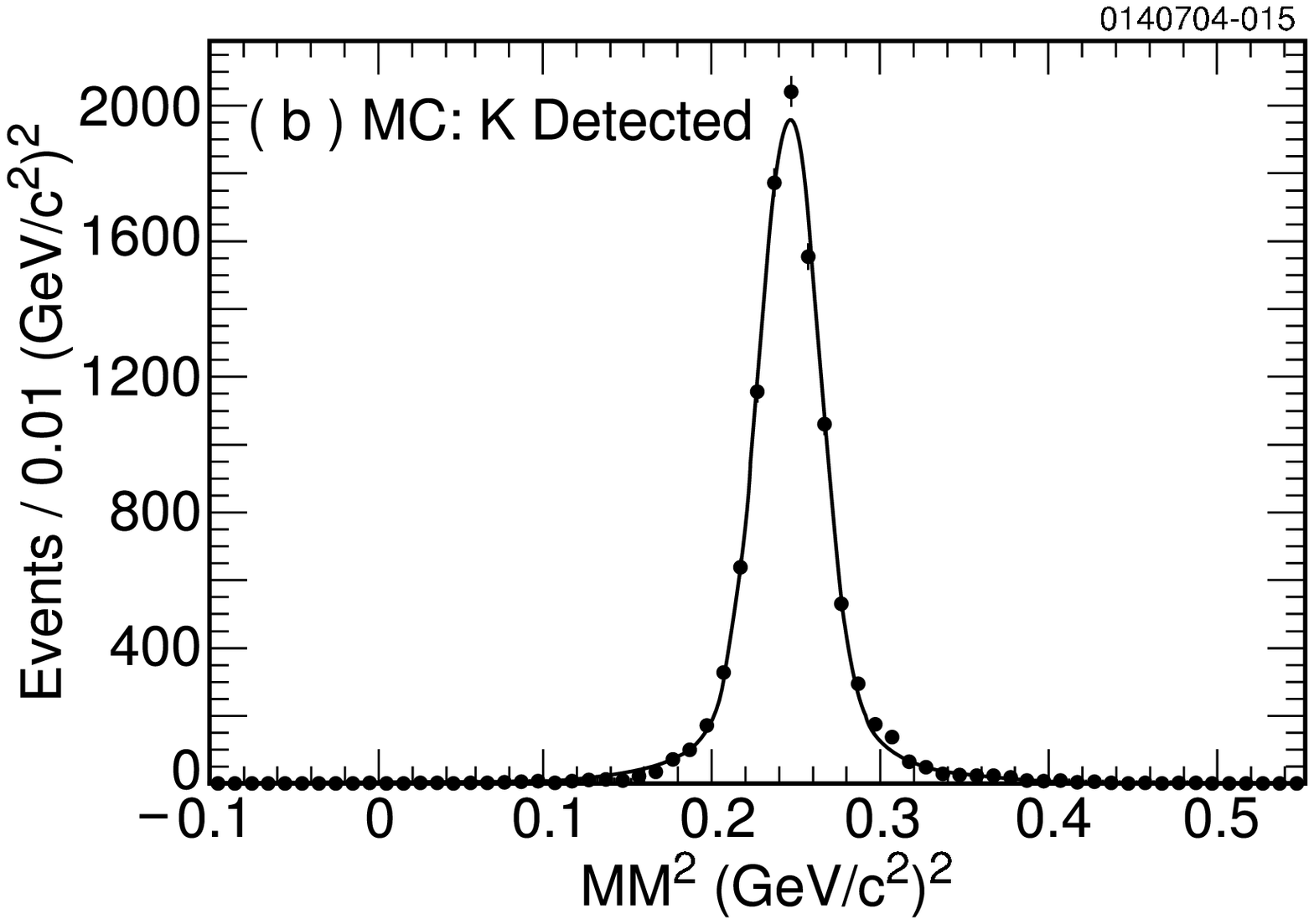}
\includegraphics[width=0.48\textwidth]{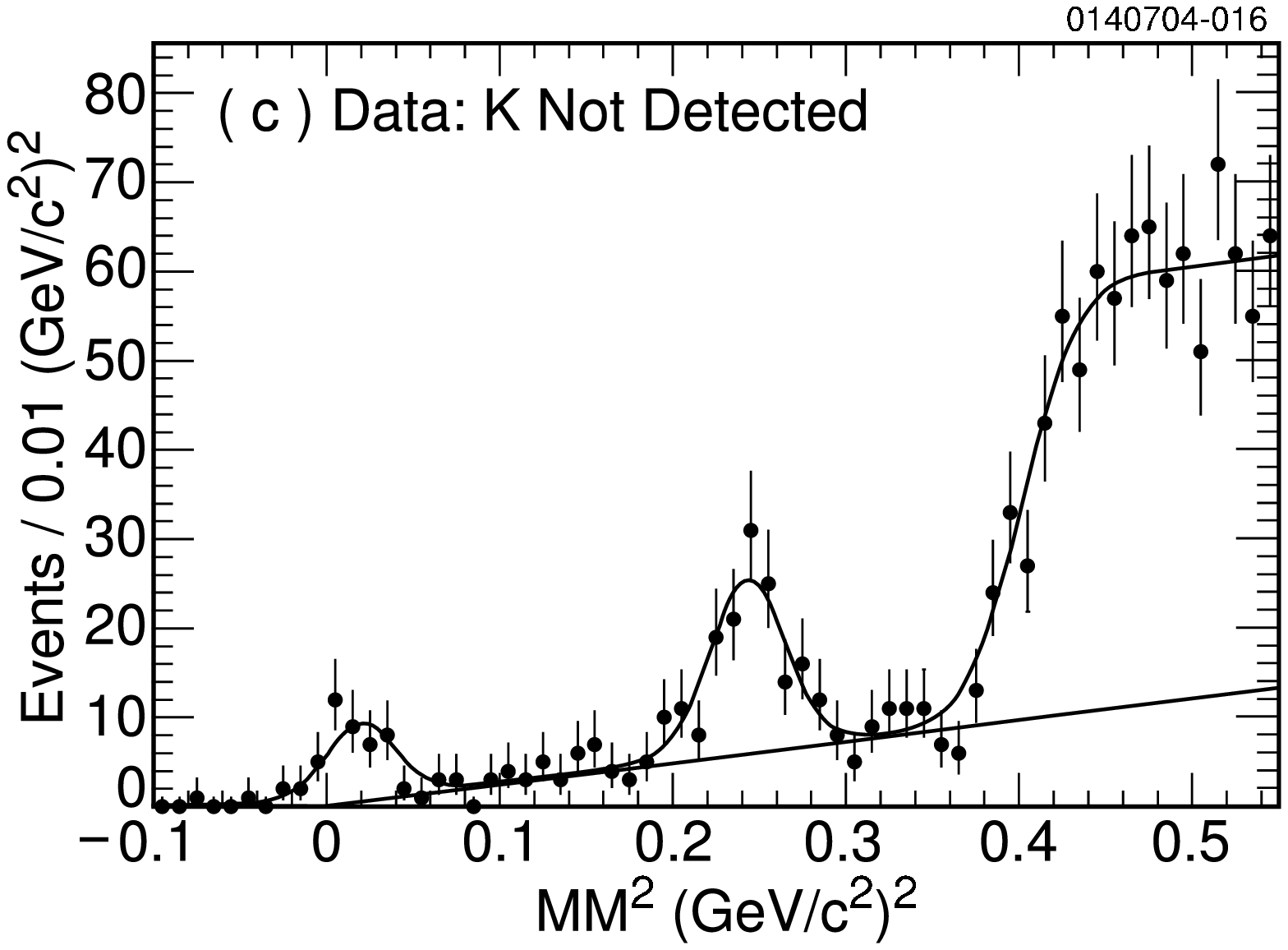}
\includegraphics[width=0.48\textwidth]{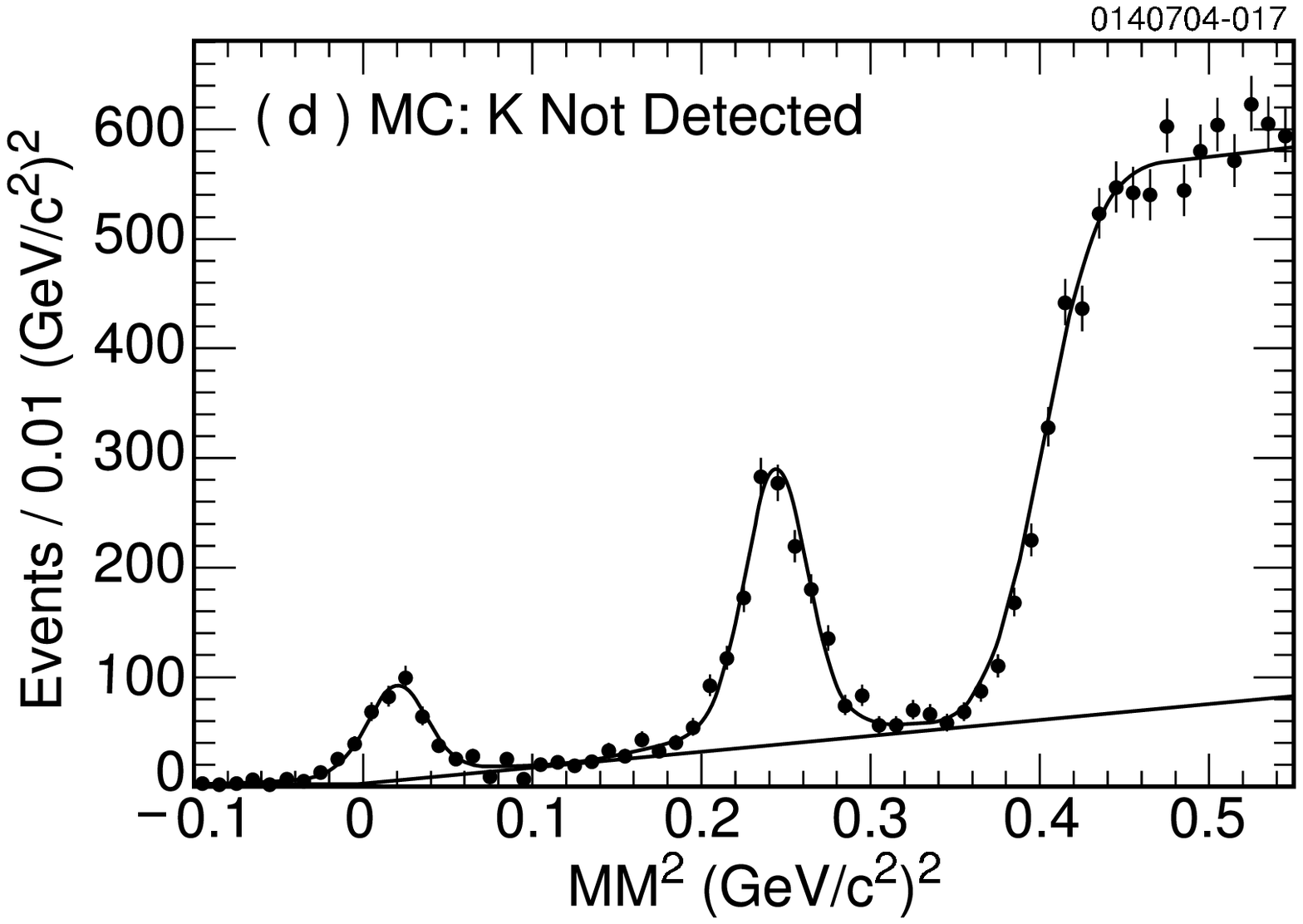}
\Endfigure{Missing mass squared distributions for measuring $\Kpm$ efficiencies
in data and Monte Carlo events in which a $\Dz$ has been found.  (a) and (b)
data and Monte Carlo events, respectively, in which the $\Kp$ from $\Dzbarkpi$
decay was found.  (c) and (d) data and Monte Carlo events, respectively, in
which the $\Kp$ from $\Dzbarkpi$ decay was not found.}{fig-keff}

\end{document}